\documentclass[journal]{IEEEtranTIE}
\usepackage{graphicx}
\usepackage{stfloats} 
\usepackage{float} 
\usepackage{subcaption} 
\usepackage{amssymb} 
\usepackage{cite}
\usepackage{picinpar}
\usepackage{amsmath}
\usepackage{url}
\usepackage{flushend}
\usepackage[latin1]{inputenc}
\usepackage{colortbl}
\usepackage{soul}
\usepackage{multirow}
\usepackage{pifont}
\usepackage{color}
\usepackage{alltt}
\usepackage[hidelinks]{hyperref}
\usepackage{enumerate}
\usepackage{siunitx}
\usepackage{breakurl}
\usepackage{epstopdf}
\usepackage{pbox}

\begin{document}
\title{ Accurate Small-Signal Modeling of Digitally Controlled Buck Converters with ADC-PWM Synchronization }

\author{
	\vskip 1em
	
	Hang Zhou, \emph{Member, IEEE},
	Yuxin Yang, \emph{Member, IEEE},
	 Branislav Hredzak, \emph{Senior Member, IEEE}\\
John Fletcher, \emph{Senior Member, IEEE}
	\thanks{This work has been submitted to the IEEE for possible publication. Copyright may be transferred without notice, after which this version may no longer be accessible.*Yuxin Yang is the corresponding author of this paper. Hang Zhou, Yuxin Yang, Branislav Hredzak and John Fletcher are with the School of Electrical Engineering \& Telecommunication University of New South Wales, NSW 2052 Australia (e-mail: Yuxin Yang: z5307358@ad.unsw.edu.au; Hang Zhou: hang.zhou@unsw.edu.au; Branislav Hredzak: b.hredzak@unsw.edu.au; John Edward Fletcher:(john.fletcher@unsw.edu.au)). }
}

\maketitle
	
\begin{abstract}
Digital control has become increasingly popular in power converters. When acquiring feedback signals such as the inductor current, synchronizing the analog-to-digital converter (ADC) with the digital pulse-width modulator (DPWM) is commonly employed to accurately track their steady-state average. However, despite its widespread use, the assumption that such synchronization has negligible impact on the small-signal behavior and loop stability is not generally valid. This paper presents an exact small-signal model for digitally controlled buck converters under constant-frequency current-mode control, explicitly accounting for DPWM-ADC synchronization. Using a sampled-data framework, the proposed model captures all sideband effects introduced by sampling, yielding precise predictions of both analog and digital loop gains, even at frequencies beyond the switching and sampling frequencies. Both asymmetrical and symmetrical carrier modulations are studied. The modified Z-transform is introduced to simplify loop gain derivation by eliminating the need for computationally intensive infinite series evaluations. A pure discrete PI parameter design method is presented, enabling low-complexity compensator design and stability assessment. This framework can be further extended to digital voltage mode control and other topologies. Experimental results verify the accuracy of the proposed model and demonstrate the impact of DPWM-ADC synchronization on loop stability.
\end{abstract}


\markboth{IEEE TRANSACTIONS ON INDUSTRIAL ELECTRONICS}%
{}

\definecolor{limegreen}{rgb}{0.2, 0.8, 0.2}
\definecolor{forestgreen}{rgb}{0.13, 0.55, 0.13}
\definecolor{greenhtml}{rgb}{0.0, 0.5, 0.0}

\section{Introduction}
\IEEEPARstart{A}{s} digital control becomes increasingly popular in power converters, its advantages in cost, programmability, and integration compare to analog designs have made it the preferred choice. \cite{Yan_Predictive} In particular, it greatly simplifies the realization of sophisticated algorithms. Moreover, it enables over-the-air (OTA) firmware updates, allowing deployed products to gain new functionality without hardware modifications. 

In common engineering practice, digital power converter design typically begins with power stage modeling using the classical state-space averaging (SSA) technique \cite{Vatche_3Port, Middlebrook}. An analog compensator is then designed in the continuous-time domain, discretized, and finally implemented in the digital controller. By averaging the piecewise linear state-spaces over one switching period, SSA eliminates the non-linearities. However, SSA neglects essential effects, such as computational delays, sampling, the presence of ripple, and sideband component coupling \cite{7869346}. Consequently, SSA loses validity even if the perturbation frequency is far below the switching frequency. While SSA leverages well-established linear time-invariant (LTI) control theory for rapid design, it fails to provide an accurate prediction of the closed-loop stability, particularly when the crossover frequency approaches the sampling frequency, leading to a potential risk of instability. 

To extend the model accuracy near or beyond the switching frequency, the sideband effect must be taken into account to form a multifrequency model\cite{Yue_Review}. As an extension to SSA, the generalized state-space averaging method (GSSA)\cite{GSSA1991, GSSA1999, 506119, GSSA2025} introduces Fourier series to account for higher-order harmonics in the time-domain waveforms. Based on the linear time-periodic (LTP) theory\cite{LTP_MIT}, the harmonic state-space (HSS)\cite{HSS_1, HSS_2} and harmonic transfer function (HTF)\cite{HTF_1, HTF_2} methods have been proposed to model frequency coupling effects. HSS employs the harmonic balance principle\cite{HSS_3} to represent an LTP system as an infinite-dimensional matrix equation set\cite{Yue_Review}, while HTF captures both direct and cross-frequency interactions. While these multifrequency modeling methods greatly enhance accuracy, they tend to produce high-order models that are computationally intensive. Consequently, computer-based tools are required, and obtaining a simple closed-form solution is often not possible.

As a compromise between model accuracy and computational complexity, researchers have proposed simplified models that include only a limited number of sidebands. Two-frequency models proposed by \cite{Qiu_1SB} and \cite{Qiu_1SB2} considered one sideband component other than the perturbed frequency $f_p$. By considering a pair of sideband components and the switching frequency component, \cite{Hsiao_1Pair} fixes the low-frequency phase error in two-frequency models. An extended frequency model that incorporates all sideband components to enhance accuracy under large ripple conditions was proposed in \cite{LiXin_AllPair}; However, the analytical form of the transfer function was not provided.

The describing function (DF) method \cite{Li2009, Li_Model} enables accurate derivation of closed-loop transfer functions and has been successfully applied to current-mode \cite{Yan_3Port} and constant on-time (COT) \cite{Tian_V2} controlled Buck converters. It has also been extended to multiphase COT converters with phase overlapping \cite{Overlap1, Overlap2}, and to passive-ripple architectures with exponentially decaying slopes \cite{Huang2024}. To simplify the design process, equivalent circuit models based on Pade approximation have been proposed in \cite{Li_Model, Yan_3Port, Tian_V2}. Despite its impressive accuracy, the DF method provides limited physical insight, does not provide open-loop characterization, and involves cumbersome, topology-specific derivations.

Building on the early sampled-data foundations \cite{Sampled_Brown, Sampled_Verghese}, a unified modeling method is proposed for various ripple-based control schemes in \cite{YanNaPartI, YanNaPartII}. By applying Shannon's sampling theorem and infinite series summation, this method rigorously captures sideband components, provides accurate predictions beyond the switching frequency, and avoids the small-ripple assumption. It also offers open-loop information with a simpler derivation than DF-based models.

Although significant progress has been made in modeling analog-controlled converters, modeling efforts dedicated to digitally controlled converters remain comparatively limited. \cite{YanNaPartII} extends the sampled-data approach to digitally controlled buck converters and derives an open-loop transfer function in the \(s\)-domain. \cite{Dragan} develops a z-domain state-space-based discrete-time model, explicitly accounting for sampling and digital-delay effects and thereby simplifying digital compensator design. Related discrete-time modeling methods have been further developed for dual active bridge converters with the effect of zero-voltage-switching transition intervals considered \cite{Costinett_DAB_2012, Costinett_DAB_2014, Costinett_DAB_Reduced_2015}, general converter analysis using discrete-time state-space modeling \cite{Baxter_Costinett_DTSS_2019}, and digitally current-mode controlled series-capacitor Buck converters (SCB) with stability-margin-based controller design \cite{Majumder_SCB_2026}. The broader role of small-signal and large-signal approaches in high-performance digital controlled DC-DC converters has also been reviewed in \cite{Kapat_Krein_Review_2020}. In addition, the nonlinear effects associated with digital PWM discretization and quantization have been investigated in \cite{Singha_DCMC_Boost_2015}.

\begin{figure}[t]
	\centering
	\includegraphics[width=\columnwidth]{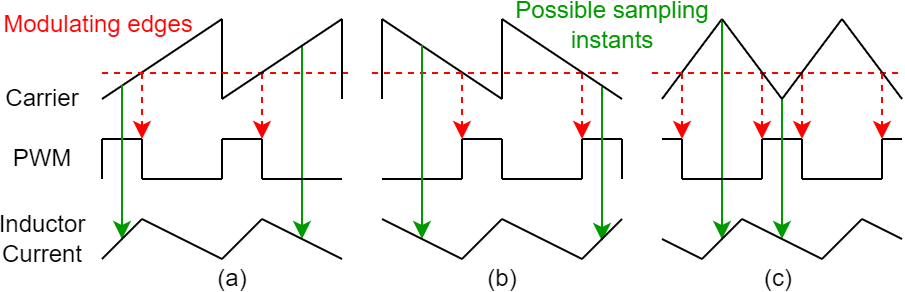}
	\caption{PWM modulation schemes with possible synchronized ADC sampling instants (vertical solid, green) and modulating edges (vertical dashed, red) highlighted. (a) TEM, up-counting sawtooth carrier. (b) LEM, down-counting sawtooth carrier. (c) Symmetrical modulation, triangular carrier. }\label{Fig_Carriers}
\end{figure}

The influence of modulation strategies on system dynamics has also been explored. Fig.\ref{Fig_Carriers} presents the modulation strategies. For example, \cite{XinLi_PWM} studies asymmetric carrier modulations and analyzes how leading-edge modulation (LEM) and trailing-edge modulation (TEM) affect loop stability and transient response in analog modulators, while \cite{Jiang} further investigates symmetric carrier modulation (SM). \cite{Ziyang_SD_PWM} compares LEM and TEM in digital pulse-width modulators (DPWM) and proposes a second-order global equivalent circuit to facilitate controller design. \cite{Bala_Digital_Boost} demonstrates that adopting LEM in a digitally controlled boost converter can eliminate the right-half-plane zero to enhance transient performance.

In addition, \cite{Lin_Accurate_Digital_Buck} and \cite{Ruan_Loopgain_Measure} show that different injection points and perturbation injection methods yield different loop gain measurements in digitally controlled converters. \cite{Ruan_Loopgain_Measure} further establishes the relationship between the responses obtained under different injection methods. Furthermore, \cite{Miaja_Subsampling} considers the case where the sampling rate is lower than the switching frequency and proposes a corresponding design methodology.

However, the impact of DPWM-ADC synchronization in digitally controlled converters has not been investigated in previous works. In practical implementations, the ADC sampling instant is often dynamically aligned with the duty cycle, such that sampling occurs at the center of the on-time or off-time interval. This allows the sampled inductor current to closely approximate its steady-state average value while minimizing the influence of noisy switching transients. However, such synchronization significantly affects the small-signal characteristics. The duty cycle perturbation affects the ADC samples. Consequently, if the sampling instant is instead assumed to be fixed at its large-signal steady-state position, the prediction will be inaccurate and can even lead to an unstable controller design. Despite its widespread use in engineering practice, to the authors' best knowledge, this effect has not yet been accounted for in any existing small-signal model.

The remainder of this paper is organized as follows.
Section~II presents the small-signal sampled-data modeling for TEM with explicit consideration of DPWM-ADC synchronization. A closed-form expression of the digital loop gain $T_{pul}$ is derived using the modified z-transform, and shows that the analog loop gain $T_i$ can be obtained from $T_{pul}$, thereby avoiding complicated infinite-series evaluations. The conclusions are also extended to LEM.
Section~III develops the small-signal model for the case of symmetrical carriers.
Section~IV compares the proposed model with simulation results and existing models. Section~V presents a PI controller design method, discusses practical considerations, and the extension of the model.
Section~VI provides experimental validation, confirming that the proposed model exhibits good agreement in all cases. Furthermore, the experiment demonstrates that introducing DPWM-ADC synchronization significantly reshapes the small-signal behavior.
Section~VII concludes the paper.

Throughout this paper, lowercase variables with a time argument (e.g. $i_L(t)$), denote time-domain signals that include both the periodic steady-state component (denoted by uppercase variables with a time argument, e.g. $I_L(t)$) and the small-signal perturbation (denoted by a hat, e.g. $\hat{i}_L(t)$). The DC value of a signal is indicated by an overline, e.g. $\overline{I_L}$.
\section{Small-Signal Modeling of Digital Buck Converters with Asymmetrical Carriers}

\begin{figure}[h]
    \includegraphics[width=\columnwidth]{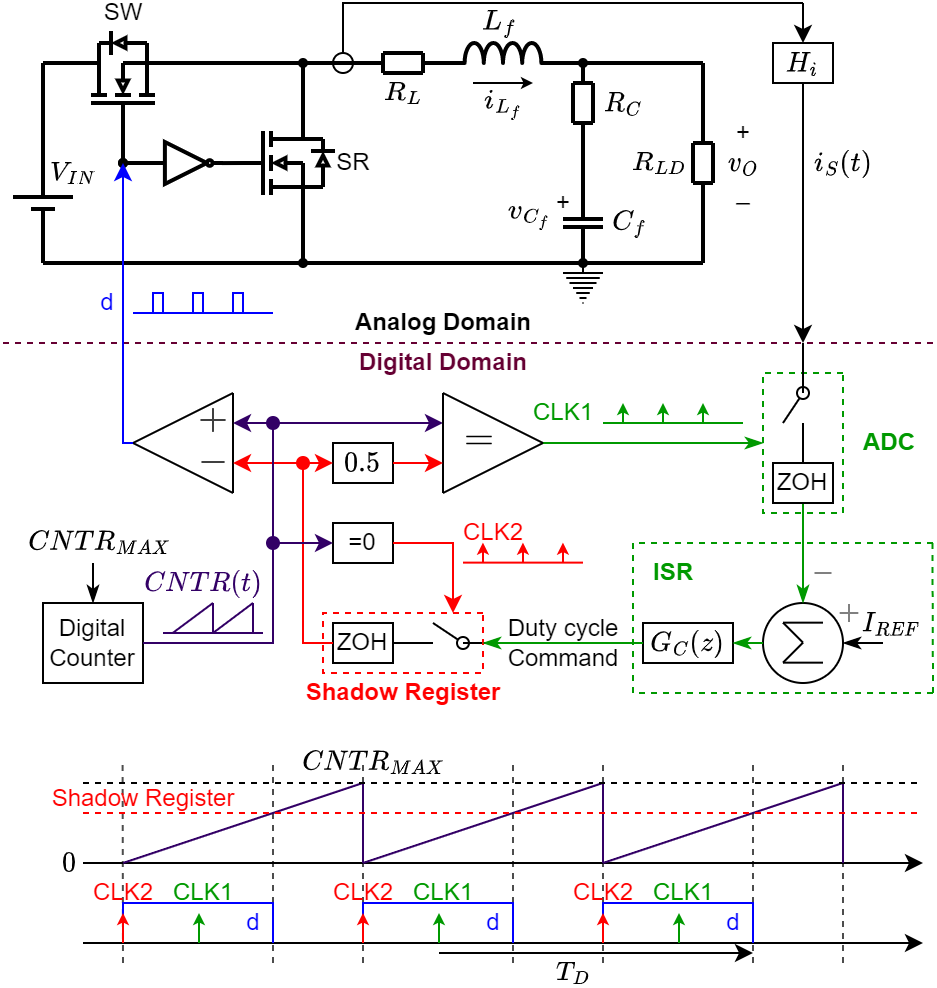}
    \caption{A digital buck with ADC sampling at the center of the PWM On-interval and synchronized to the TEM DPWM.}
    \label{fig:TEM_Buck}
\end{figure}

Fig. \ref{fig:TEM_Buck} illustrates a digitally controlled synchronous Buck converter employing current-mode control and operating in forced continuous conduction mode (FCCM). The digital counter generates an up-counting sawtooth carrier $CNTR(t)$ with an amplitude of $CNTR_{MAX}$. When $CNTR$ reaches zero, the counter updates the shadow register. The shadow register serves to suppress spurious PWM pulses during abrupt changes in the duty-cycle command. The DPWM output will be high if the shadow register output is higher than $CNTR$ and will be low elsewhere. This configuration implements a TEM DPWM, in which the PWM rising edge is fixed at the beginning of each cycle, while the falling edge varies and moves according to the duty-cycle command.

\begin{figure*}[t]
	\centering
	\includegraphics[width=0.9\textwidth]{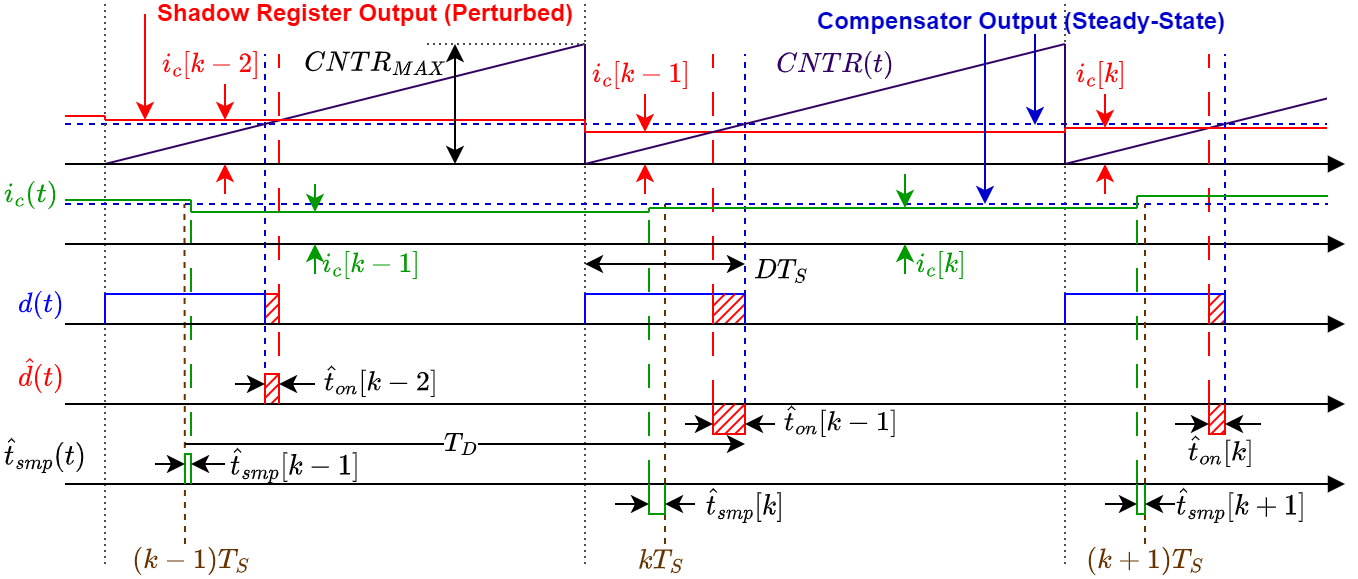}
    \caption{Discrete-domain waveforms of a TEM buck converter. Dense and sparse vertical dashed lines denote the sampling instants at steady state and under perturbation, respectively. The ADC samples at integer multiples of $T_S$.}
    \label{fig:TEM_Buck_BlockDiagram_Timing}
\end{figure*}

To simplify the analysis, we define the beginning of each cycle as the steady-state ADC sampling instant ($kT_S$). The block diagram and timing are presented in Fig.\ref{fig:TEM_Buck} and Fig.\ref{fig:TEM_Buck_BlockDiagram_Timing}, where $H_i$ is the current sensor gain and $T_s$ is the switching period. In practice, $H_i$ also includes the ADC gain. The scaled inductor current is denoted as $i_s$. The duty-cycle to the inductor current transfer function and the digital PI compensator transfer function\cite{YanNaPartII} are:

\begin{equation}
    G_{id}(s) = V_{IN}\bigg[sL_f+R_L+(\frac{1}{sC_f}+R_C) \parallel R_{LD}\bigg]^{-1} 
    \label{eq:gid}
\end{equation}

\begin{equation}
    G_C(z) = K_P+\frac{K_iT_s}{1-z^{-1}}
    \label{eq:digital_pi}
\end{equation}

The ADC and shadow register are modeled as two sets of sample-and-hold blocks. Under the small-signal assumption, the DPWM comparator behaves as another ideal sampler, allowing the duty-cycle perturbation 
$\hat{d}$ to be treated as an impulse train \cite{YanNaPartI}. When a sampling-hold-sampling sequence operates at the same sampling rate, it can be equivalently represented by a single sampler followed by a delay equal to the time difference between the two sampling instants. By applying this lemma twice, the 3 ideal samplers and 2 ZOH blocks found above can be simplified to a single sampler followed by a delay $T_D$, without any ZOH. Under the small-signal assumption, $T_D$ is defined as the steady-state time interval between the ADC sampling instant and the moment when the new calculated duty-cycle command takes effect.

Since DPWM-ADC synchronization guaranties that sampling occurs exactly at the midpoint of the PWM on-interval, as shown in Fig.\ref{fig:TEM_Buck_IL}, in each cycle, the ADC sampling instant $t_{smp}$ is related to the on-time perturbation $\hat{t}_{on}$ of the previous cycle:

\begin{figure}[h]
    \centering
    \includegraphics[width=0.9\columnwidth]{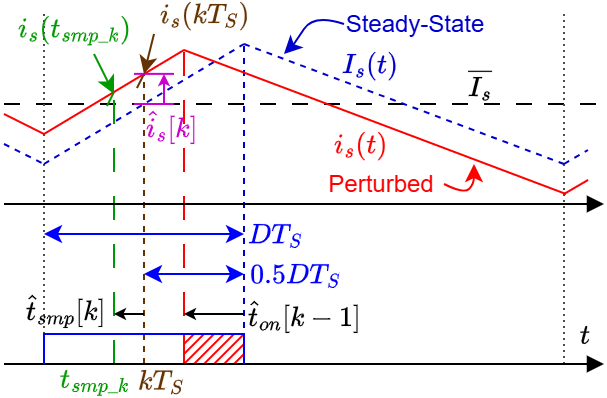}
    \caption{Sampling instant variation due to DPWM-ADC synchronization. Dense and sparse vertical dashed lines represent sampling points under steady-state and perturbed, respectively.}
    \label{fig:TEM_Buck_IL}
\end{figure}

\begin{equation}
    t_{smp\_k} = kT_S+0.5\hat{t}_{on}[k-1]
    \label{eq:t_smp_k}
\end{equation}

Under the small-signal assumption, the k-th sampled inductor current $i_L(t_{smp\_k})$ can be approximated its 1st-order Taylor expansion as:

\begin{equation}
    i_s(t_{smp\_k}) \approx i_s(kT_S)+\frac{di_s(t)}{dt}\bigg|_{t=kT_S} \times \frac{\hat{t}_{on}[k-1]}{2}
    \label{eq:is_smp_k}
\end{equation}
where $i_s(kT_S)$ is the scaled inductor current sampled at fixed instants $kT_S$, $I_s(kT_S)$ is the steady-state sampled inductor current, which equals the average inductor current $\overline{I_L}$:
\begin{equation}
    i_s(kT_S) = i_s[k] = I_s(kT_S) + \hat{i}_s[k] = \overline{I_L}H_i+\hat{i}_s[k]
    \label{eq:is_kts}
\end{equation}
Extracting perturbation terms from \eqref{eq:is_smp_k} and \eqref{eq:is_kts} yields:
\begin{equation}
    \hat{i}_{smp}[k] = \hat{i}_s[k] + \frac{di_s(t)}{dt}\bigg|_{t=kT_S} \times \frac{\hat{t}_{on}[k-1]}{2}
    \label{eq:i_smp_hat}
\end{equation}

\begin{figure}[t]
    \centering
    \includegraphics[width=\columnwidth]{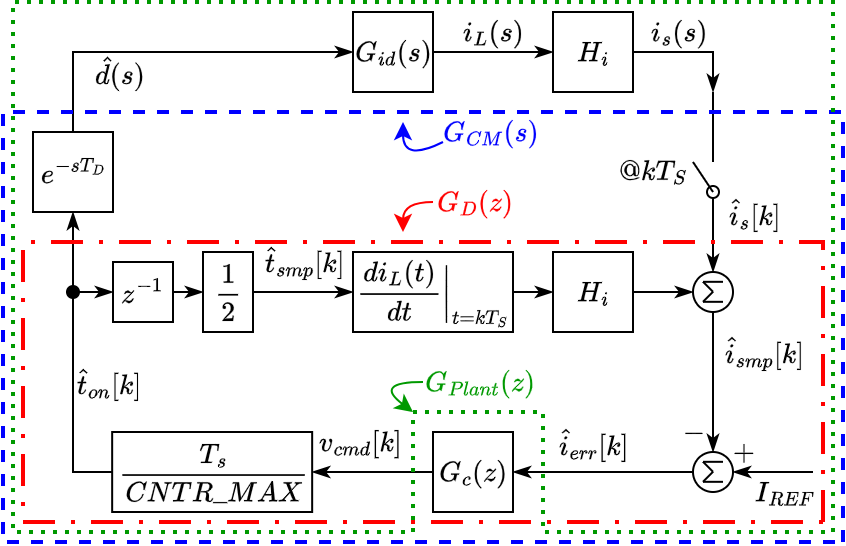}
    \caption{Block diagram of TEM digital buck without ZOH.}
    \label{fig:TEM_Buck_BlockDiagram2}
\end{figure}

Therefore, with \eqref{eq:i_smp_hat} and Lemma 1, Fig.\ref{fig:TEM_Buck} can be simplified to Fig.\ref{fig:TEM_Buck_BlockDiagram2}. $G_{CM}(s)$ represents the analog feedback-to-modulator output transfer function and accounts for the side-band effects due to sampling, the "CM" subscript stands for controller and modulator. $G_{Plant}(z)$ denotes the plant seen by the digital PI compensator $G_C(z)$. The transfer function of the pure discrete part in Fig. \ref{fig:TEM_Buck_BlockDiagram2}, from $\hat{i}_s[k]$ to $\hat{t}_{on}[k]$, can be derived as:

\begin{equation}
G_D(z) = \frac{-G_C(z)\frac{T_S}{CNTR_{MAX}}}{1-[-G_C(z)\frac{T_S}{CNTR_{MAX}}]\times H_{sync}(z)}
\end{equation}
where the feedthrough due to DPWM-ADC alignment is:
\begin{equation}
    H_{sync}(z) = \frac{z^{-1}}{2}\frac{di_L(t)}{dt}\bigg|_{t=kT_S} \times H_i
    \label{eq:H_sync}
\end{equation}

According to the sampling theorem, the impulse train $\hat{i}_s[k]$'s s-domain representation $\hat{i}_s^*(s)$ is:
\begin{equation}
    \hat{i}_s^*(s) = \hat{i}_s^*(s+jn\omega_S) = \frac{1}{T_s} \sum_{n=-\infty}^{\infty} \hat{i}_s\!\left(s + j n \omega_s\right)
    \label{eq:is_sampling}
\end{equation}
isolating the $n=0$ case of the infinite summation gives:
\begin{equation}
    \hat{i}_s^*(s) = \frac1{T_s}\hat{i}_s(s) + \frac1{T_s}\sum_{\substack{n=-\infty \\ n \neq 0}}^{\infty} \hat{i}_s\!\left(s + j n \omega_s\right)
    \label{eq:is_sampling_expanded}
\end{equation}
The second term of \eqref{eq:is_sampling_expanded} models sideband couplings. From Fig.\ref{fig:TEM_Buck_BlockDiagram2}, $\hat{i}_s(s)$ can be expressed as:
\begin{equation}
    \hat{i}_s(s) = \hat{t}_{on}^*(s)e^{-sT_D}G_{id}(s)H_i
    \label{eq:i_s_continuous}
\end{equation}
Since $\hat{t}_{on}[k]$ is also an impulse train, its s-domain representations are:
\begin{equation}
    \hat{t}_{on}^*(s) = \hat{t}_{on}^*(s+jn\omega_S) = G_D(e^{sT_S})\hat{i}_s^*(s)
    \label{eq:d_hat_discrete}
\end{equation}
Expanding the infinite summation term in \eqref{eq:is_sampling_expanded} with \eqref{eq:i_s_continuous} and then replacing $\hat{t}_{on}^*(s+jn\omega_S)$ with $\hat{t}_{on}^*(s)$ according to \eqref{eq:d_hat_discrete} gives:
\begin{equation}
    \begin{aligned}
    \hat{i}_s^*(s) = \frac{1}{T_s}&\bigg[\hat{i}_s(s)+\hat{t}_{on}^*(s)\times\\
    &\sum_{\substack{n=-\infty \\ n \neq 0}}^{\infty}e^{-(s+jn\omega_S)T_D}G_{id}(s+jn\omega_S)H_i\bigg]
    \end{aligned}
    \label{eq:is_star_expanded}
\end{equation}
Substituting \eqref{eq:is_star_expanded} into \eqref{eq:d_hat_discrete} to eliminate $\hat{i}_s^*(s)$, then isolating $\hat{t}_{on}^*(s)$ and $i_s(s)$ gives:
\begin{equation}
    \begin{aligned}
    &\frac{\hat{t}_{on}^*(s)}{\hat{i}_s(s)} = \frac{\frac{G_D(z)}{T_S}}
    {1 - \frac{G_D(z)}{T_S} 
      \displaystyle\sum_{\substack{n=-\infty \\ n \neq 0}}^{\infty}
      e^{-(s+jn\omega_S)T_D}G_{id}(s+jn\omega_S)H_i}
    \\&\quad\quad\quad\text{where: } \quad z=\exp(sT_S)
    \end{aligned}
    \label{eq:is_to_d}
\end{equation}

Hence, $G_{CM}(s)$ can be derived from \eqref{eq:is_to_d} as:
\begin{equation}
    G_{CM}(s) \triangleq -\frac{\hat{d}(s)}{\hat{i}_s(s)} = -e^{-sT_D}\frac{\hat{t}_{on}^*(s)}{\hat{i}_s(s)}
\end{equation}

The analog loop gain $T_i$ can be expressed as:

\begin{equation}
    T_i(s) = G_{CM}(s)G_{id}(s)H_i
\end{equation}

From Fig. \ref{fig:TEM_Buck_BlockDiagram2}, the plant transfer function seen by the digital compensator can be easily found as:

\begin{equation}
    G_{Plant}(z) = \frac{T_S}{CNTR_{MAX}} \bigg[H_{sync}(z)+G_{MZ}(z) \bigg]
    \label{eq:G_Plant}
\end{equation}
where:
\begin{equation}
    G_{MZ}(z)=\frac1{T_S}\displaystyle\sum_{\substack{n=-\infty}}^{\infty}
      \big[e^{-sT_D}G_{id}(s)H_i\big]_{s->s+jn\omega_S}
    \label{eq:G_MZ}
\end{equation}

The digital loop gain $T_{pul}$ is then:
\begin{equation}
    T_{pul}(z) = G_{Plant}(z) G_C(z)
    \label{eq:T_pul}
\end{equation}

The loop gain measurement reads differently as the injection point changes \cite{Lin_Accurate_Digital_Buck, Ruan_Loopgain_Measure}. The relationship between $T_i$ and $T_{pul}$ is found to be:

\begin{equation}
    T_i(s) = \frac{T_0(s)}{1+T_{pul}(\exp(sT_S))-T_0(s)}
    \label{eq:ruan}
\end{equation}
where:
\begin{equation}
    T_0(s) = G_c(e^{sT_S})\frac{1}{CNTR_{MAX}}e^{-sT_D}G_{id}(s)H_i
\end{equation}

\eqref{eq:ruan} agrees with the conclusion of \cite{Ruan_Loopgain_Measure} and provides a convenient method to evaluate the analog loop gain from the digital loop gain. Furthermore, \eqref{eq:G_MZ} resembles the modified Z-transform\cite{jury_ztransform} of $G_{id}H_i$ with a sampling delay of $T_D$. One can perform partial fraction on $G_{id}H_i$ and then utilize the following modified Z-transform identity\cite{yang_ztransform} to evaluate \eqref{eq:G_MZ}:

\begin{equation}
    \mathcal{Z}_m\!\left\{\sum_{r=1}^{\mathrm{PoleCount}}\frac{n_r}{s+d_r},\,T_p\right\}
    =\sum_{r=1}^{\mathrm{PoleCount}} \frac{n_r\,e^{d_r T_p}}{z\,e^{d_r T_S}-1}.
    \label{eq:Zm}
\end{equation}
where the delay $T_p$ must satisfy $T_p\in (0,\,T_S)$, i.e., strictly within one sampling period.  
If the desired sampling delay $T_D$ is not in this range, it can be decomposed as:
\begin{equation}
    T_D = kT_S + T_p, \quad k \in \mathbb{Z}, \; T_p \in (0,\,T_S).
\end{equation}
In this case, the modified $z$-transform of a delay $T_D$ can be obtained by multiplying $z^{-k}$ with the modified $z$-transform of a delay $T_p$:
\begin{equation}
    G_{MZ}(z)= \mathcal{Z}_m\{G_{id}H_i,\,T_D\}=z^{-k}\mathcal{Z}_m\{G_{id},\,T_p\}H_i.
    \label{eq:G_MZ_Calc}    
\end{equation}
where in the case of TEM, $k=1$ and $T_p=0.5DT_S$, as shown in Fig.\ref{fig:TEM_Buck} and Fig.\ref{fig:TEM_Buck_BlockDiagram_Timing}.

The above analysis for TEM can be extended to LEM. 
Table~\ref{tab:sampling_delay} summarizes the effective sampling delay $T_D$ 
under different modulation modes and sampling strategies.

\begin{table}[]
    \centering
    \caption{$T_D$ Under Various Asymmetrical Modulations and Different Sampling Positions}
    \label{tab:sampling_delay}
    \begin{tabular}{l l l l}
        \hline
        Modulation + Sampling Position & k & $T_p$ \\
        \hline
        TEM, Sample at the on-interval center   & $1$ & $0.5DT_S$ \\
        TEM, Sample at the off-interval center  & $0$ & $0.5(1+D)T_S$ \\
        LEM, Sample at the on-interval center   & $0$ & $(1-0.5D)T_S$ \\
        LEM, Sample at the off-interval center  & $1$ &  $0.5(1-D)T_S$ \\
        \hline
    \end{tabular}
\end{table}

By applying \eqref{eq:G_MZ_Calc} to \eqref{eq:G_Plant},
the plant transfer function seen by the digital compensator 
can be obtained for both TEM and LEM cases, which provides the basis for loop compensation design, e.g., using the pole-placement method.
\section{Small-Signal Modeling of Digital Buck Converters with Symmetrical Carriers}

\begin{figure*}[t]
	\centering
	\includegraphics[width=0.9\textwidth]{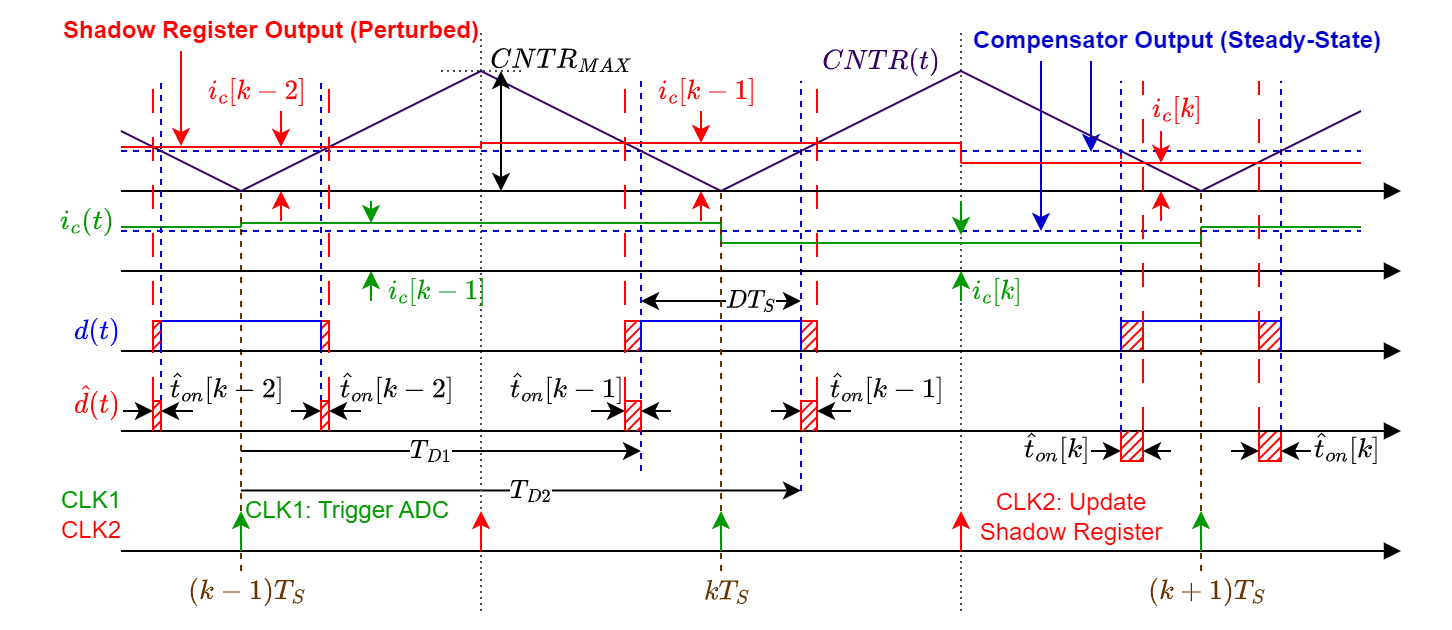}
    \caption{Key discrete-domain waveforms of a buck converter with a symmetrical carrier. Dense and sparse vertical dashed lines denote the sampling instants at steady state and under perturbation, respectively. The ADC samples at integer multiples of $T_S$.}
    \label{fig:TEM_Buck_BlockDiagram_Timing_Sym}
\end{figure*}

The above analysis for asymmetrical carriers can be extended to the case where the carrier is a symmetric triangular waveform. Fig.\ref{fig:TEM_Buck_BlockDiagram_Timing_Sym} illustrates the key waveforms of a digital buck converter with a triangular carrier. The ADC samples when $CNTR$ reaches 0, which corresponds to the center of the PWM on-interval. Afterwards, the new duty cycle command is loaded into the shadow register when $CNTR$ reaches $CNTR_{MAX}$. Compared to the asymmetrical case, two important differences arise. First, in the symmetrical case, the ADC sampling instant is always stationary with respect to the carrier and does not vary with any perturbation. Consequently, there will be no $H_{\text{sync}}(z)$. Second, each impulse $\hat{t}_{on}[k]$ now maps to two impulses in $\hat{d}$ that have different delays ($T_{D1}$ and $T_{D2}$), which is similar to the sampled-data COT model in~\cite{YanNaPartII}.

In this case, the plant transfer function in the $z$-domain seen by the digital compensator is given by:
\begin{equation}
    \begin{aligned}
    G_{\text{Plant,SYM}}(z) = \frac{T_s}{2\,\text{CNTR}_{\max}} \bigg(&\mathcal{Z}_m \{G_{id}, T_{D1}\} + \\
    &\mathcal{Z}_m \{G_{id}, T_{D2}\} \bigg) H_i
    \end{aligned}
    \label{eq:G_plant_sym}
\end{equation}

where $T_{D1}$ and $T_{D2}$ are given in Table \ref{tab:sampling_delay_sym}. This table also accounts for the alternative configuration where sampling occurs at the off-interval center, i.e., when $CNTR = \text{CNTR}_{\max}$ and the shadow register is updated at $CNTR=0$.

\begin{table}[]
    \centering
    \caption{$T_{D1}$ and $T_{D2}$ Under Symmetrical Modulation and Different Sampling Positions}
    \label{tab:sampling_delay_sym}
    \begin{tabular}{l l l}
        \hline
        Sampling Position & $T_{D1}$ & $T_{D2}-T_{D1}$ \\
        \hline
        Sample at the on-interval center   & $(1-0.5D)T_S$ & $DT_S$ \\
        Sample at the off-interval center  & $0.5(1+D)T_S$ & $(1-D)T_S$ \\
        \hline
    \end{tabular}
\end{table}

Equation \eqref{eq:G_plant_sym} provides the plant transfer function for digital compensator design.  
The open-loop transfer function in the digital domain can then be computed from \eqref{eq:T_pul}, while the corresponding analog-domain loop gain can still be obtained directly using \eqref{eq:ruan}. In this case, $T_0(s)$ is modified as:
\begin{equation}
    T_{0,\text{SYM}}(s) = \frac{G_c(e^{sT_S})}{2\text{CNTR}_{\max}}
    (e^{-sT_{D1}}+e^{-sT_{D2}})G_{id}(s)H_i
\end{equation}

\section{Comparison Between Existing Small-Signal
 Models and the Proposed Analytical Model}
 
For a fair comparison, all models are evaluated under the same set of circuit and control parameters, summarized in Table~\ref{tab:parameters}. The exact form of \eqref{eq:gid} is used throughout, rather than the weak-coupling approximation adopted in~\cite{YanNaPartII}.

\begin{table}[h]
    \centering
    \caption{Parameters of the Buck Converter}
    \label{tab:parameters}
    \begin{tabular}{l l l}
        \hline
        Symbol & Description & Value \\
        \hline
        $V_{\text{IN}}$ & Input voltage & $12V$ \\
        $I_\text{REF}$ & Output current set-point (scaled by $H_i$) & $0.2$ \\
        $T_S$ & Switching period & $10\mu s$ \\
        $L_f$ & Inductance & $4.1\mu$H \\
        $R_L$ & DC winding resistance (DCR) of $L_f$ & $51m\Omega$ \\
        $C_f$ & Output capacitance & $404\mu$F \\
        $R_C$ & Equivalent series resistance (ESR) of $C_f$ & $32m\Omega$ \\
        $R_{LD}$ & Load resistance & $1.4\Omega$ \\
        $CNTR_{MAX}$ & Digital counter upper bound & $2000$ \\
        $\omega_c$ & Target angular crossover frequency & $10\text{kHz}$ \\
        $\theta_\text{PM}$ & Target angular phase margin & $45^\circ$ \\
        $Kp$ & Compensator proportional gain & $200.4$ \\
        $KiT_S$ & Compensator integration gain & $175.38$ \\
        \hline
    \end{tabular}
\end{table}

\begin{figure*}[!b]
    \centering
    \makebox[\textwidth][c]{%
        \subfloat[]{
            \includegraphics[width=0.34\textwidth]{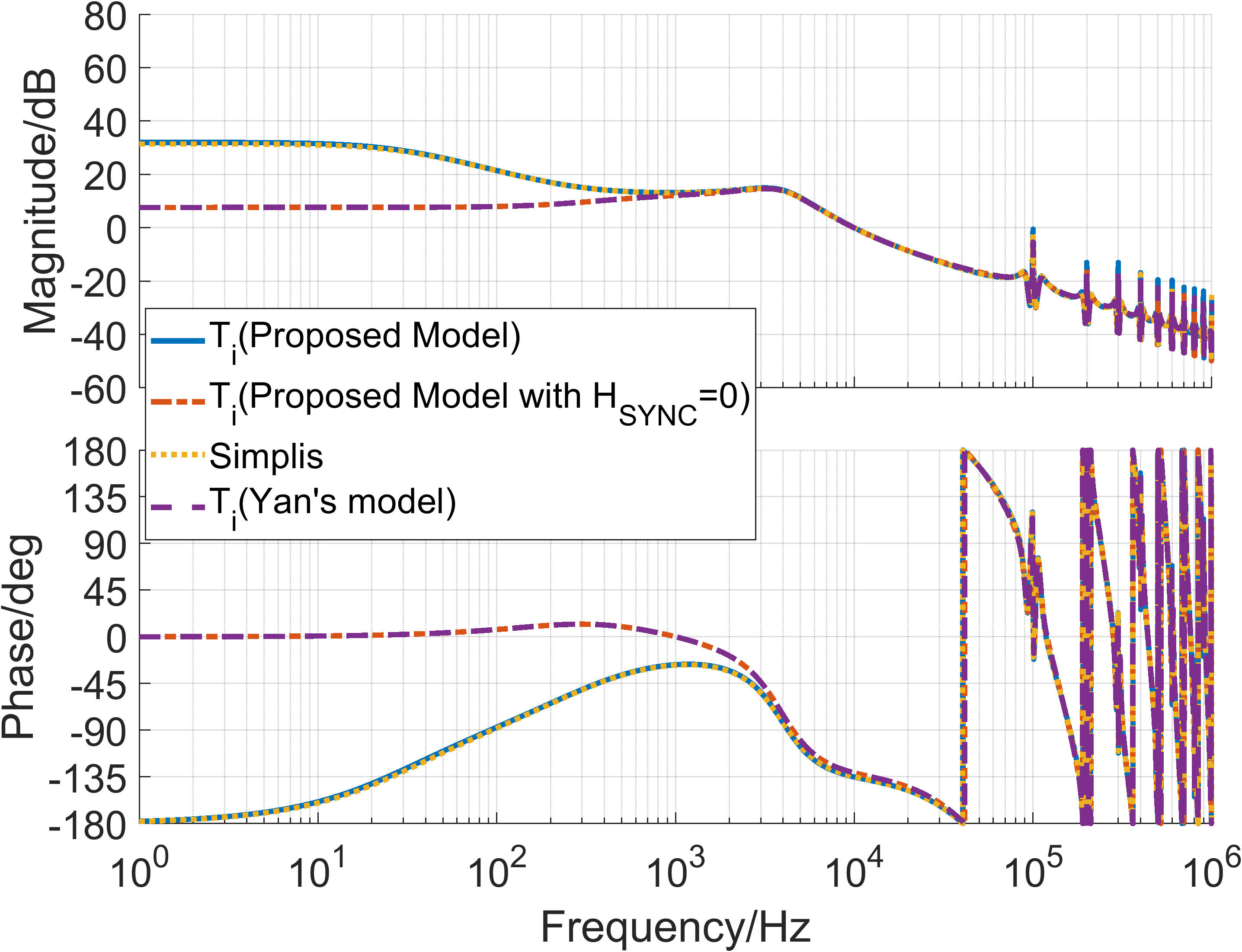}
            \label{fig:TEM_Off_Ti}
        }\hspace{-0.03\textwidth}
        \subfloat[]{
            \includegraphics[width=0.34\textwidth]{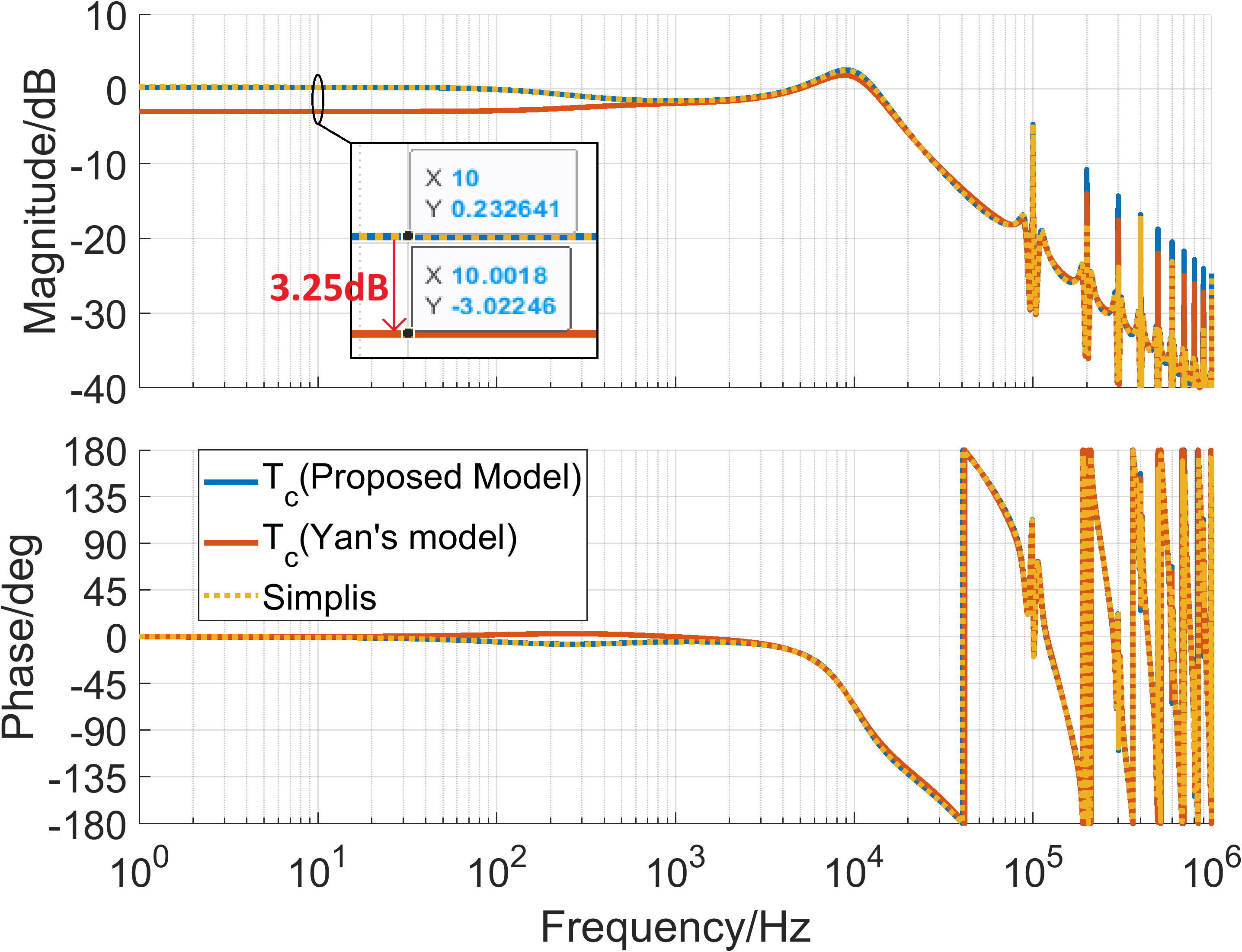}
            \label{fig:TEM_Off_Tc}
        }\hspace{-0.03\textwidth}
        \subfloat[]{
            \includegraphics[width=0.34\textwidth]{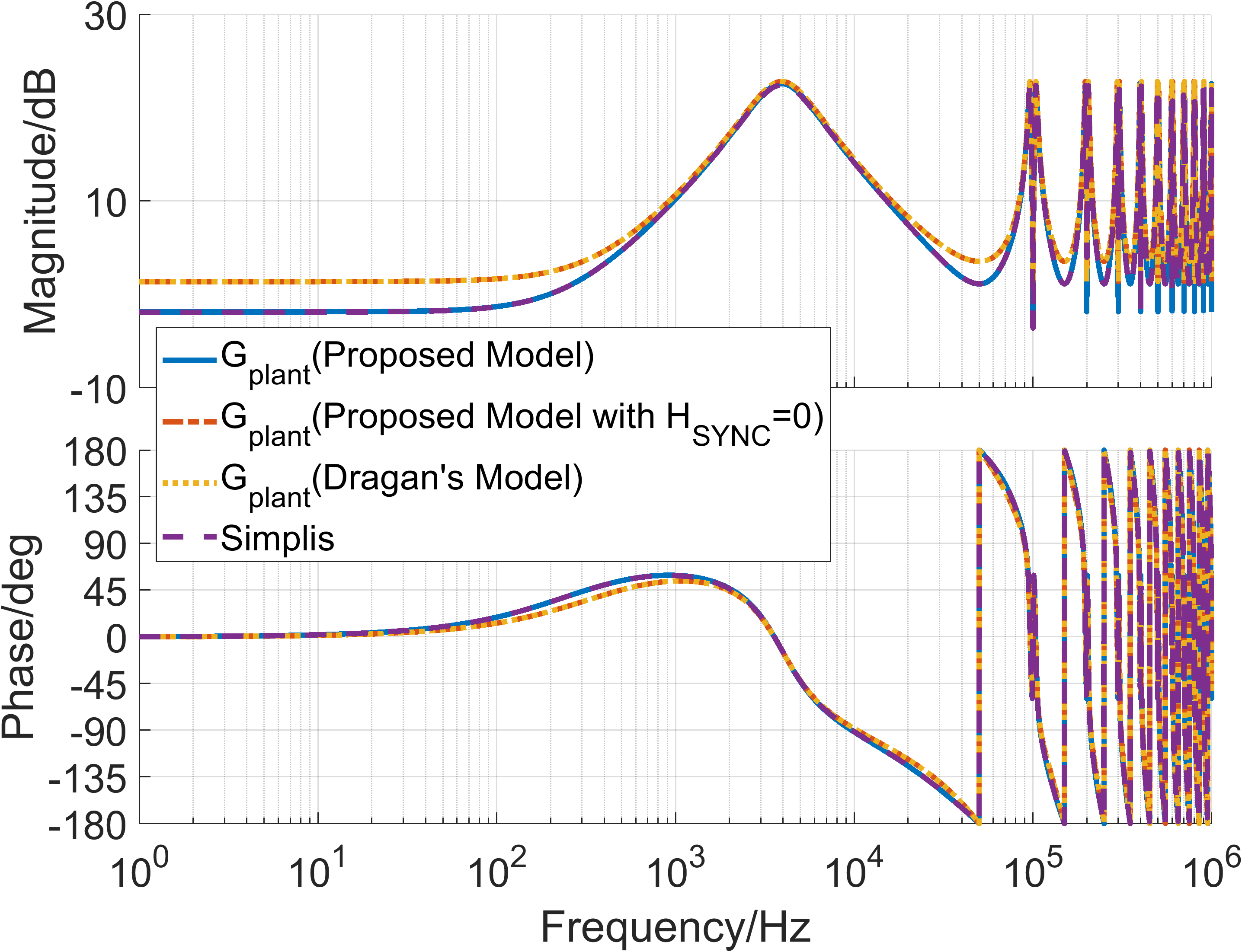}
            \label{fig:TEM_Off_Gplant}
        }
    }

    \caption{Comparison of analytical and simulated Bode diagrams under TEM modulation with turn-off-centered sampling.
    (a) Analog loop gain: proposed model, Yan's model~\cite{YanNaPartII}, and simulation.
    (b) Analog closed-loop transfer function: proposed model, Yan's model~\cite{YanNaPartII}, and simulation.
    (c) Plant transfer function seen by the digital compensator: proposed model, Dragan's model~\cite{Dragan}, and simulation.}
    \label{fig:TEM_Off_Ti_Tc_Gplant}
\end{figure*}

Fig.~\ref{fig:TEM_Off_Ti} compares the analog loop gain predicted by the proposed model, Yan's model~\cite{YanNaPartII}, and SIMPLIS simulation. The results were obtained under TEM modulation with turn-off-centered sampling. Since the proposed model explicitly incorporates the DPWM-ADC synchronization effect through the $H_{\text{sync}}(z)$ term, its prediction shows excellent agreement with the simulation results. When the $H_{\text{sync}}(z)$ path is removed, i.e., without DPWM-ADC synchronization, $T_i$ given by the proposed model reduces to Yan's model~\cite{YanNaPartII}. In contrast, Yan's model~\cite{YanNaPartII} deviates by around 25~dB lower, translating to significant inaccuracies in the predicted closed-loop gain($T_c(s)=T_i(s)/(1+T_i(s))$~\cite{YanNaPartI}), as depicted in Fig.~\ref{fig:TEM_Off_Tc}. The 3.25-dB difference in $T_c(s)$ can lead to approximately a 30\% error in the predicted steady-state average tracking value. This observation further confirms that, in a digitally controlled power converter, without proper DPWM-ADC synchronization, the steady-state error remains large even if the digital compensator contains an integral term.

Fig.\ref{fig:TEM_Off_Gplant} presents the plant transfer function as perceived by the digital compensator, predicted by the proposed model, Dragan's purely discrete model~\cite{Dragan}, and SIMPLIS simulations. It is observed that Dragan's model, which neglects DPWM-ADC synchronization, consistently deviates from the simulation results. In contrast, the proposed model achieves an excellent match across the entire frequency range, validating its superior accuracy. When the $H_{\text{sync}}(z)$ path is removed, $T_{pul}$ given by the proposed model reduces to Dragan's model~\cite{Dragan}.

\begin{figure}[t]
    \centering
    \subfloat[]{
        \includegraphics[width=0.85\columnwidth]{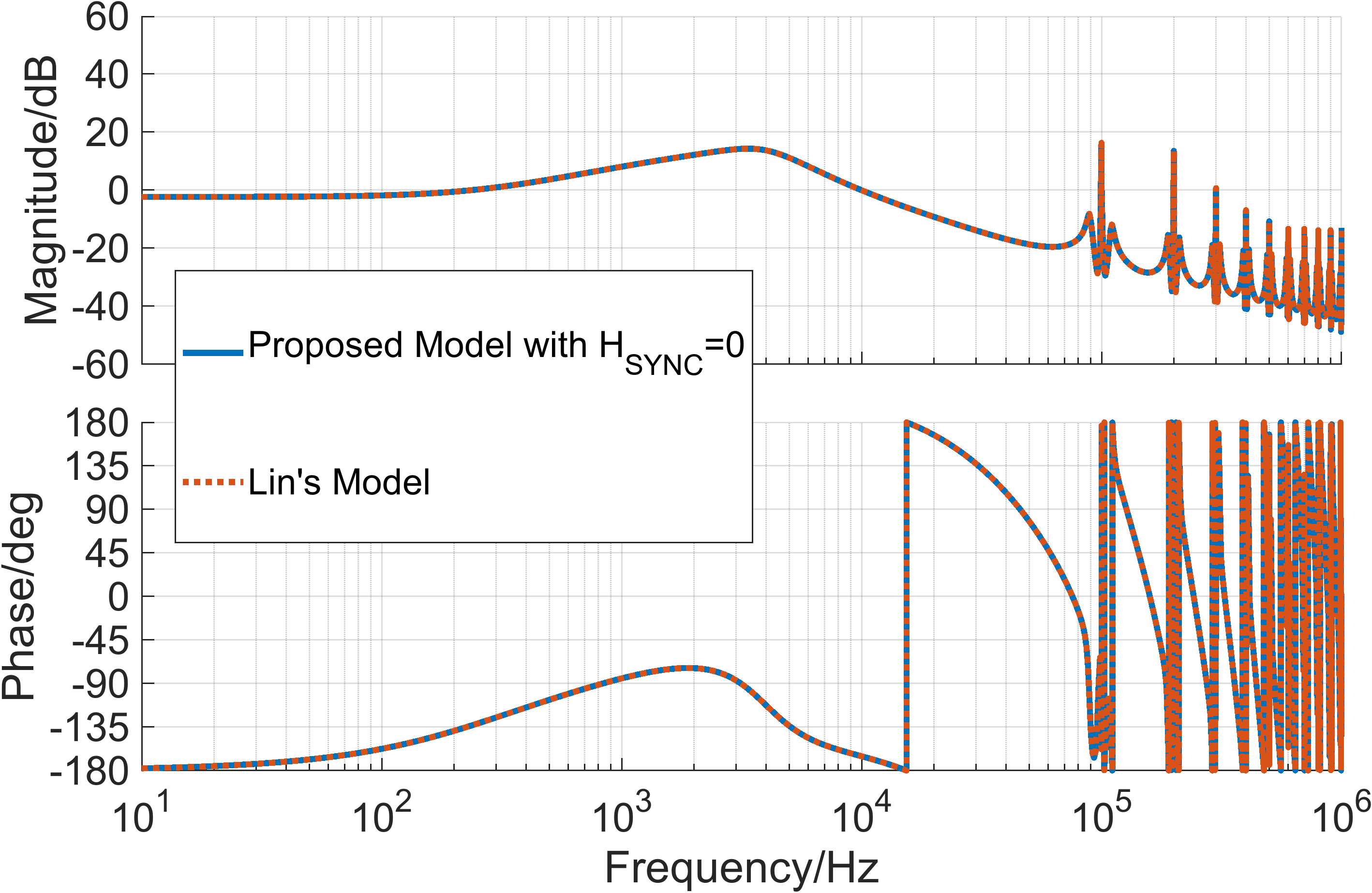}
        \label{fig:Ti_Lin}
    }
    \\[0.5em]
    \subfloat[]{
        \includegraphics[width=0.85\columnwidth]{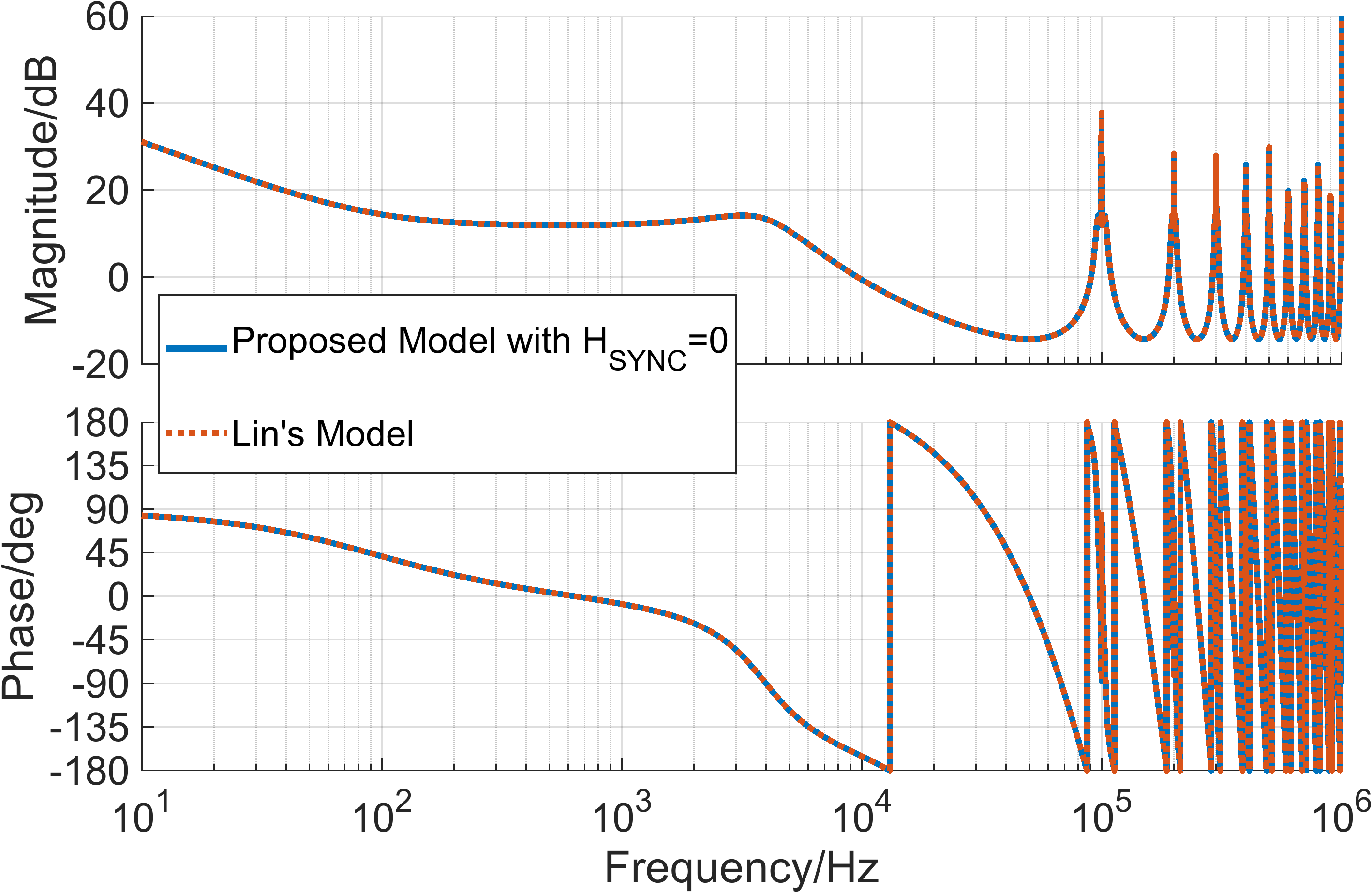}
        \label{fig:Tpul_Lin}
    }
    \caption{Comparison of the proposed model with the $H_{sync}$ path removed (set to 0) and Lin's model~\cite{Lin_Accurate_Digital_Buck}.
    (a) $T_i$ (correspond to $T_I$ in ~\cite{Lin_Accurate_Digital_Buck}).
    (b) $T_{pul}$ (correspond to $T_{II}$ in ~\cite{Lin_Accurate_Digital_Buck}).}
    \label{fig:Ti_Tpul_Lin}
\end{figure}

Lin's model~\cite{Lin_Accurate_Digital_Buck} does not include an explicit counterpart of $T_D$. Moreover, the model is formulated specifically for TEM, and the ADC always samples at the PWM rising edge, with the updated duty command taking effect at the PWM falling edge in the following switching period. Therefore, when interpreted within the timing framework used in this paper, Lin's model corresponds to an equivalent fixed $T_D$ of $(1+D)T_S$. Fig.\ref{fig:Ti_Tpul_Lin} presents a comparison between Lin's model and the proposed model with $H_{sync}(z)$ path removed.

\begin{figure}[h]
    \centering
    \includegraphics[width=0.85\columnwidth]{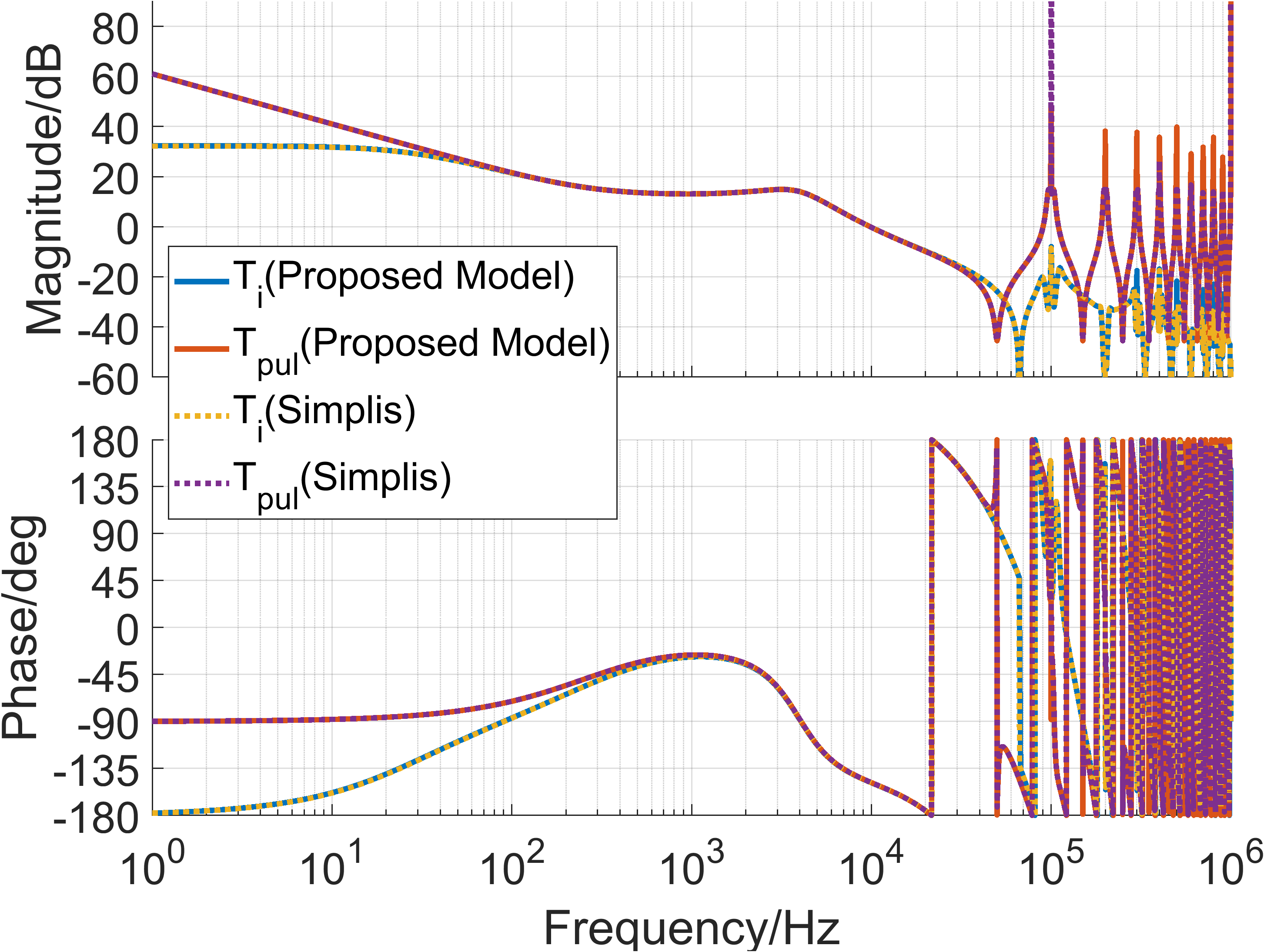}
    \caption{Analog and digital loop gains under symmetrical triangular modulation with turn-off centered sampling: comparison of the proposed model and simulation.}
    \label{fig:SYM_Off_TiTpul}
\end{figure}

To further verify the proposed model, Fig.~\ref{fig:SYM_Off_TiTpul} compares its predictions with SIMPLIS simulations under symmetrical modulation with turn-off-centered sampling. Both $T_i$ and $T_{pul}$ show good agreement with the simulations, although they are different from each other because they are obtained using different perturbation methods. As reported in~\cite{Lin_Accurate_Digital_Buck,Ruan_Loopgain_Measure}, $T_i$ and $T_{pul}$ exhibit close agreement in the mid-frequency range (100~Hz to half the switching frequency), but significant discrepancies appear at low and high frequencies. At low frequencies, $T_{pul}$ behaves as an ideal integrator, theoretically yielding infinite DC gain, whereas the $T_i$ settles to a finite value. This distinction can be rigorously verified by evaluating the limits of \eqref{eq:ruan} and \eqref{eq:T_pul} as $s \to 0$. This result also highlights the implication that the zero steady-state error perceived by the digital compensator only indicates that the ADC sample follows the reference, and does not necessarily imply that the analog variable being controlled fully tracks the reference.

Table~\ref{tab:comparison} compares the proposed model with several established modeling approaches, including the sampled-data model in \cite{YanNaPartII}, the discrete-time model in \cite{Dragan}, and the HTF-based model in \cite{Lin_Accurate_Digital_Buck}. Unlike the existing methods, the proposed model simultaneously incorporates DPWM-ADC synchronization, supports all three DPWM carrier types, and remains computationally inexpensive with low derivation complexity. In addition, letting $H_{sync}(z)=0$ can reduce the proposed model to the models in \cite{YanNaPartII,Dragan,Lin_Accurate_Digital_Buck}.

\begin{table}[t]
    \centering
    \caption{Comparison between the proposed model and existing models of digital DC-DC converters}
    \label{tab:comparison}
    \begin{tabular}{l c c c c}
        \hline
        Model & \begin{tabular}{@{}c@{}}Consider \\ DPWM-ADC \\ Alignment\end{tabular} & Perspective & Modulation & Complexity \\
        \hline
        Yan's\cite{YanNaPartII} & No & Analog & TEM & Moderate \\
        Dragan's\cite{Dragan} & No & Digital & TEM & Low \\
        Lin's\cite{Lin_Accurate_Digital_Buck} & No & Analog & TEM & High \\
        This work & Yes & \begin{tabular}{@{}c@{}}Analog \& \\ Digital\end{tabular} &  \begin{tabular}{@{}c@{}}TEM, LEM, \\ Symmetrical\end{tabular} & Low \\
        \hline
    \end{tabular}
\end{table}

\section{Application of the Proposed Model}
\subsection{PI Controller Design}
To demonstrate the usefulness of the proposed model, this section presents an example of a digital PI controller entirely in the \(z\)-domain without relying on \(s\)-domain approximations. Given the well-known design targets, the angular crossover frequency $\omega_c$ and the angular phase margin $\theta_\text{PM}$ must satisfy:

\begin{equation}
G_{\mathrm{plant}}G_{\mathrm{C}}\left(e^{j\theta_c}\right)
=
e^{j(\theta_\text{PM}-\pi)}
\quad\text{, where }\theta_c = \omega_c T_S
\label{eq:jury}
\end{equation}

Therefore, the PI parameters for an asymmetrically modulated digitally controlled Buck converter can be found as:
\begin{equation}
K_P
=
W_R
+
W_I
\tan\left(\frac{\theta_c}{2}\right)
\text{, }
K_i
=
-\frac{2}{T_S}
W_I
\tan\left(\frac{\theta_c}{2}\right)
\end{equation}

where $Sn = \frac{di_L(t)}{dt}\bigg|_{t=kT_S}$, the other variables are:

\begin{equation}
W_R
=
-\frac{CNTR_{MAX}}{T_S}
\frac{
P_R\cos(\theta_{PM})+P_I\sin(\theta_{PM})
}{
P_R^2+P_I^2
}
\end{equation}
\begin{equation}
W_I
=
\frac{CNTR_{MAX}}{T_S}
\frac{
P_I\cos(\theta_{PM})-P_R\sin(\theta_{PM})
}{
P_R^2+P_I^2
}
\end{equation}

\begin{equation}
P_R =
\frac{S_n H_i}{2}\cos\theta_c
+
H_i
\left[
M_R\cos(k\theta_c)
+
M_I\sin(k\theta_c)
\right]
\end{equation}

\begin{equation}
P_I =
-\frac{S_n H_i}{2}\sin\theta_c
+
H_i
\left[
M_I\cos(k\theta_c)
-
M_R\sin(k\theta_c)
\right]
\end{equation}

\begin{equation}
M_R
=
e^{\alpha T_p}
\left(
M_{R,-}+M_{R,+}
\right)
\text{,  }
M_I
=
e^{\alpha T_p}
\left(
M_{I,-}+M_{I,+}
\right)
\end{equation}

\begin{equation}
M_{R,\pm}
=
\frac{
U\left[\cos\left(\theta_c\pm\beta T_S\right)-e^{-\alpha T_S}\right]
\pm
V\sin\left(\theta_c\pm\beta T_S\right)
}{
e^{\alpha T_S}
-
2\cos\left(\theta_c\pm\beta T_S\right)
+
e^{-\alpha T_S}
}
\end{equation}

\begin{equation}
M_{I,\pm}
=
\frac{
\pm V\left[\cos\left(\theta_c\pm\beta T_S\right)-e^{-\alpha T_S}\right]
-
U\sin\left(\theta_c\pm\beta T_S\right)
}{
e^{\alpha T_S}
-
2\cos\left(\theta_c\pm\beta T_S\right)
+
e^{-\alpha T_S}
}
\end{equation}

\begin{equation}
U=p\cos(\beta T_p)+q\sin(\beta T_p)
\end{equation}

\begin{equation}
V=p\sin(\beta T_p)-q\cos(\beta T_p)
\end{equation}

\begin{equation}
\alpha=\frac{a_1}{2a_2},\qquad \beta=\frac{\sqrt{4a_2a_0-a_1^2}}{2a_2}
\end{equation}

\begin{equation}
p=\frac{V_{IN}}{2L_f},\qquad q=\frac{V_{IN}(a_1-2L_f)}{2L_f\sqrt{4a_2a_0-a_1^2}}
\end{equation}

\begin{equation}
a_2 = C_f L_f (R_C + R_{LD}),\qquad a_0 = R_L + R_{LD}
\end{equation}

\begin{equation}
a_1 = L_f + C_f \left(R_C R_L + R_C R_{LD} + R_L R_{LD}\right)
\end{equation}

As shown in Fig.~\ref{fig:TEM_Off_Ti}, the resulting PI parameters satisfy the prescribed design objectives in Table~\ref{tab:parameters} for TEM with turn-off-centered sampling. This design procedure is further experimentally validated in the following section. It should also be noted that \eqref{eq:jury} is also applicable to SM. The corresponding derivation is omitted here due to page limits.

\subsection{Implementation Considerations}
\begin{figure}[t]
    \centering
    \subfloat[]{
        \includegraphics[width=0.85\columnwidth]{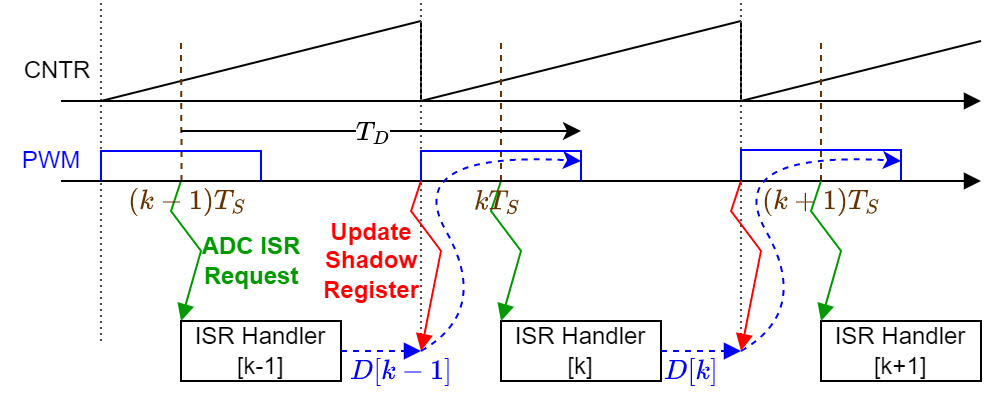}
        \label{fig:TEM_On_Timing}
    }
    \\[0.5em]
    \subfloat[]{
        \includegraphics[width=0.85\columnwidth]{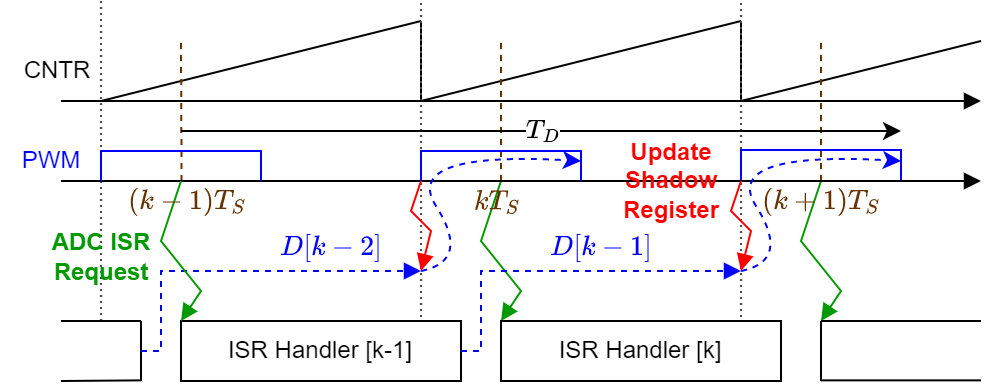}
        \label{fig:TEM_On_Timing_Overrun}
    }
    \caption{Timing of TEM with turn-on-centered sampling under different ISR execution times.
    (a) Normal case, where the ISR completes before the next shadow register update.
    (b) Overrun case, where the ISR misses the shadow register update and introduces an additional one-cycle delay.}
    \label{fig:TEM_On_Timing_ISR}
\end{figure}

In a microcontroller-based (MCU) implementation, the end of ADC conversion triggers an interrupt service routine (ISR) that runs the PI algorithm to set a new duty ratio. Taking TEM with turn-on-centered sampling as an example, Fig.~\ref{fig:TEM_On_Timing} shows the normal timing case, where the ISR execution time is less than $(1-0.5D)T_S$, the new duty ratio takes effect in the next switching cycle, and $T_D$ is consistent with Table~\ref{tab:sampling_delay}. However, if the ISR execution time is between $(1-0.5D)T_S$ and $T_S$, as illustrated in Fig.~\ref{fig:TEM_On_Timing_Overrun}, it will add a delay of $T_S$ to $T_D$. If the ISR execution time exceeds $T_S$, the digital controller cannot be implemented at the specified switching frequency.

As discussed in~\cite{Kapat_Krein_Review_2020}, a large $T_D$ can degrade the dynamic response. Tables~\ref{tab:sampling_delay} and~\ref{tab:sampling_delay_sym} can be used to select modulation and sampling combinations with smaller $T_D$, which are generally more favorable for high-bandwidth control design. However, a smaller $T_D$ also reduces the available ISR execution time for the ISR, which may require a faster MCU. The difference between TEM and SM is further reflected by comparing Fig.~\ref{fig:TEM_Off_Ti} with Fig.~\ref{fig:SYM_Off_TiTpul}. Their crossover frequencies are similar, but the additional delay $T_{D2}$ associated with SM significantly reduces the phase margin from $45^\circ$ to around $30^\circ$.

\begin{figure}[t]
    \centering
    \includegraphics[width=0.9\columnwidth]{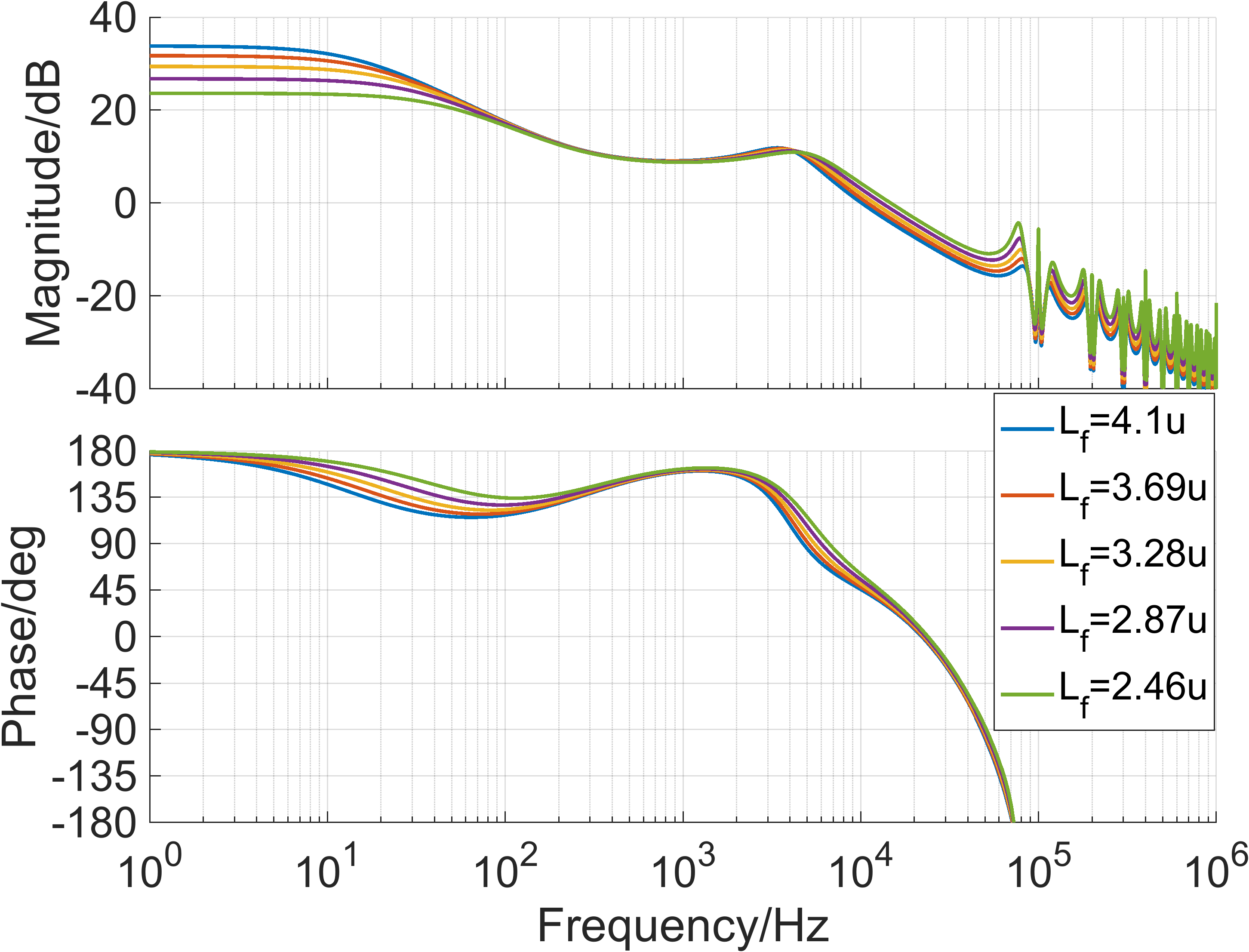}
    \caption{$T_i$, as $L_f$ is swept from $4.1\mu H$ to $2.46\mu H$.}
    \label{fig:LfScan}
\end{figure}

In practice, $L_f$ may saturate as $i_{L_f}$ increases, thereby reducing its effective inductance. To evaluate its impact on the loop stability, $L_f$ is swept from its nominal value to 40\% below the nominal value, as shown in Fig.~\ref{fig:LfScan}, which indicates that $\omega_c$ increases from $2\pi\times10~\mathrm{kHz}$ to $2\pi\times14.74~\mathrm{kHz}$, while the phase margin decreases from $45^\circ$ to $36^\circ$. On the other hand, the aging-induced increases in $C_f$ and ESR have a relatively minor impact on high-$\omega_c$ designs. Moreover, a larger ESR tends to stabilize the loop. For constant-current operation, the PI parameters should be designed for the maximum $V_{IN}$, where $G_{id}$ reaches its maximum DC gain.

\subsection{Extension of the Proposed Model}
In the derivation presented in Section~II, the current sensor is assumed to have sufficiently high bandwidth and is therefore represented by a constant gain $H_i$, which holds true for most designs. For cost-sensitive designs employing a low-bandwidth current-sense amplifier (with a transfer function of $G_i(s)$), the proposed model remains applicable. In this case, $G_i(s)$ should be included with $G_{id}(s)$ as the modified $Z$-transform input in \eqref{eq:G_MZ_Calc}. In addition, the slope term in $H_\text{sync}$ \eqref{eq:H_sync} should be modified accordingly to reflect the scaled slope at $t=kT_S$.

The proposed framework can also be extended to voltage-mode control by replacing $G_{id}$ in \eqref{eq:G_MZ_Calc} with the corresponding $s$-domain duty-to-output-voltage transfer function, namely $G_{vd}$\cite{YanNaPartI}. This extension is also applicable to other topologies with corresponding $G_{id}$ and $G_{vd}$ used, such as Boost converters\cite{YanNaPartII} and SCBs\cite{Majumder_SCB_2026}.

\section{Experimental Verification}
\begin{figure}[b]
    \centering
    \includegraphics[width=0.9\columnwidth]{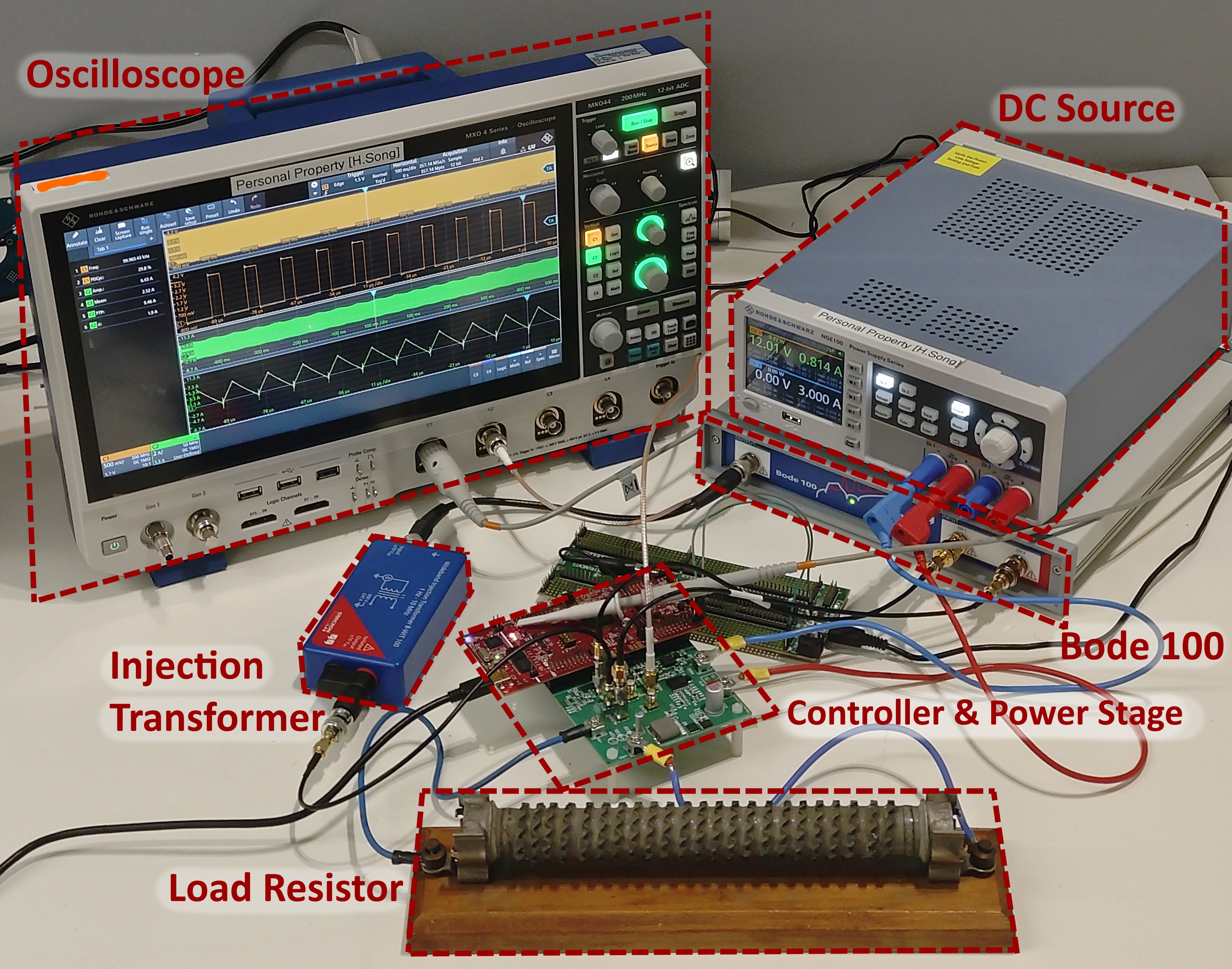}
    \caption{Experiment setup for loop gain measurements.}
    \label{fig:setup}
\end{figure}

To further validate the proposed model, an FCCM Buck converter was built as shown in Fig.~\ref{fig:setup}. A floating-point DSP (TMS320F28379) clocked at 200 MHz runs the control law. Both the switching frequency and ADC sampling rate are set to 100 kHz. A high-resolution PWM is employed so that its resolution far exceeds the 12-bit ADC resolution, thereby avoiding limit cycles caused by duty-cycle quantization. Because the resolutions of DPWM and ADC are sufficiently high, quantization effects\cite{Singha_DCMC_Boost_2015} are negligible.

\begin{figure*}[!t]
    \centering
    \makebox[\textwidth][c]{%
        \subfloat[]{
            \includegraphics[width=0.34\textwidth]{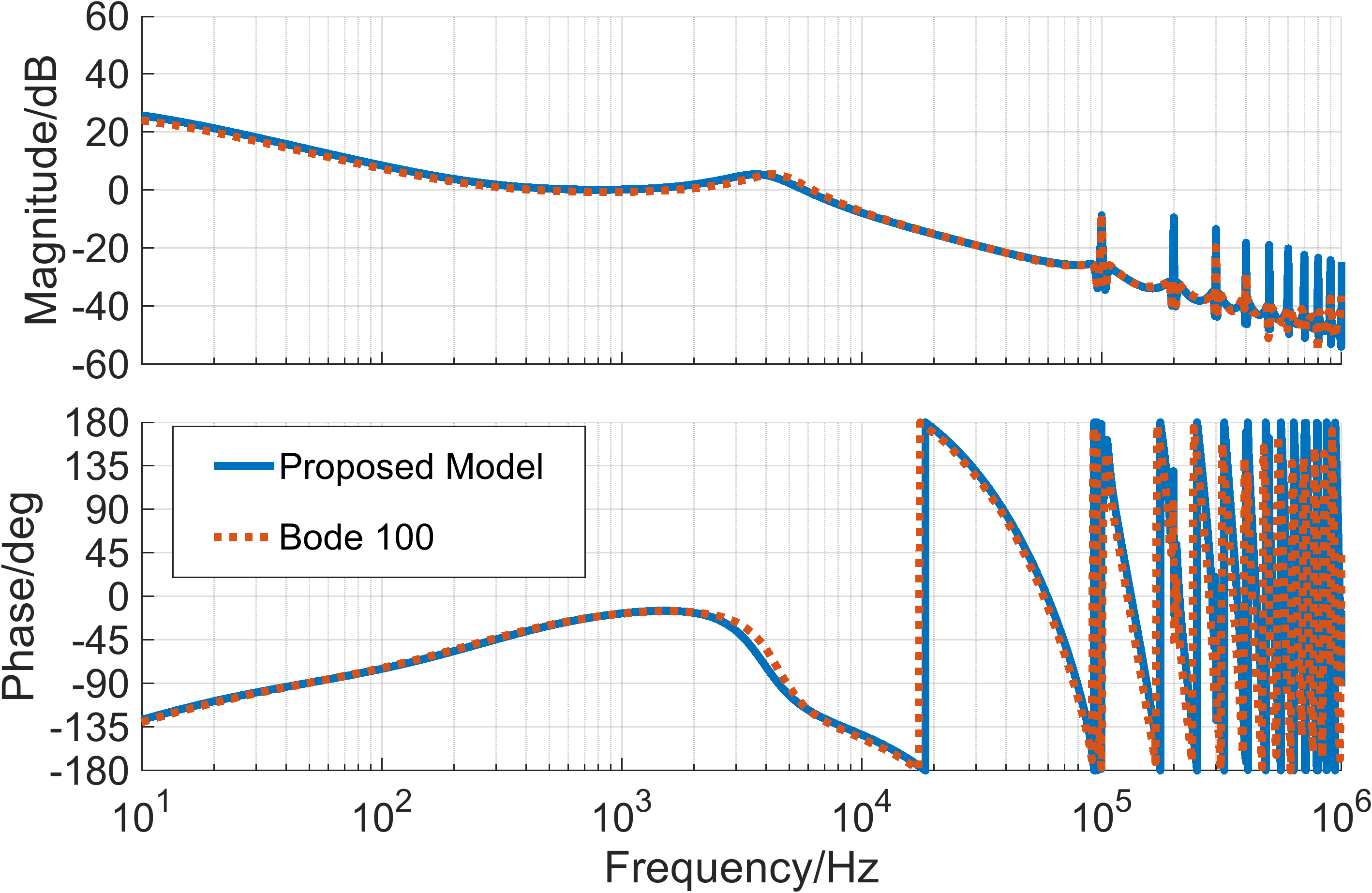}
            \label{fig:TEM_On_Ti_5V_2A}
        }\hspace{-0.03\textwidth}
        \subfloat[]{
            \includegraphics[width=0.34\textwidth]{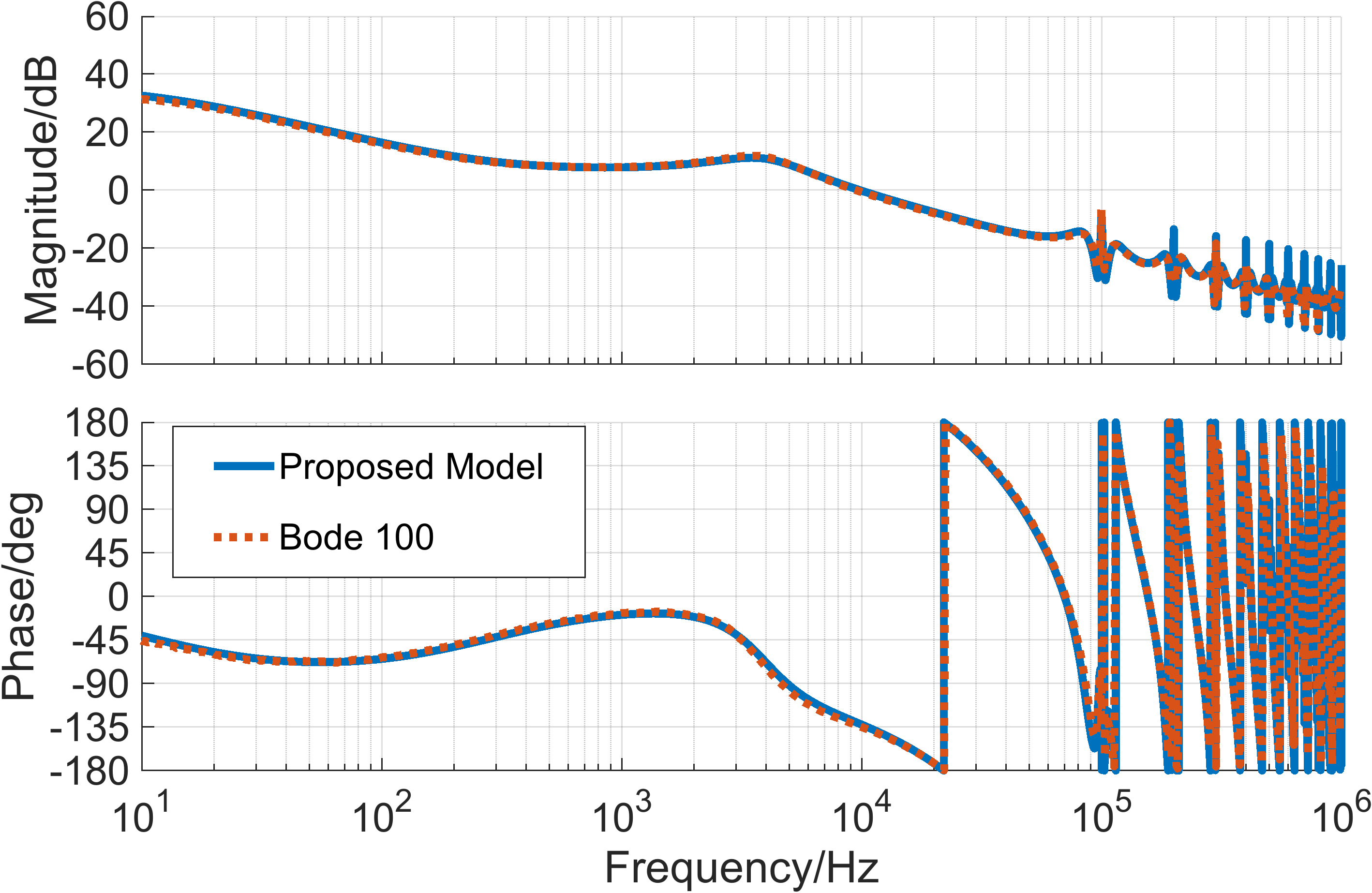}
            \label{fig:TEM_On_Ti_12V_2A}
        }\hspace{-0.03\textwidth}
        \subfloat[]{
            \includegraphics[width=0.34\textwidth]{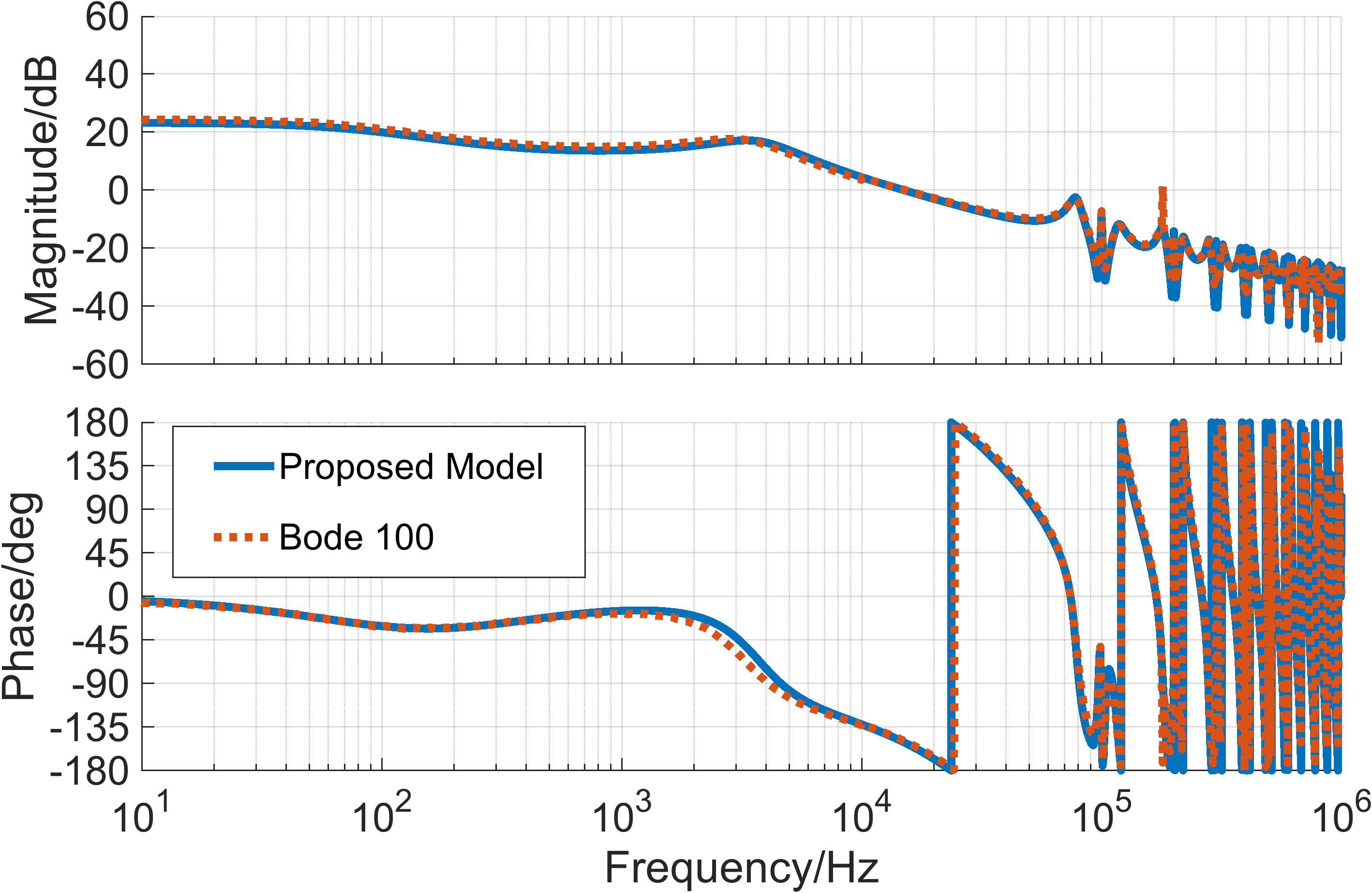}
            \label{fig:TEM_On_Ti_24V_2A}
        }
    }

    \caption{Bode diagram of the calculated and measured analog loop gain $T_i$, 
    (a) $V_{\mathrm{IN}}=5\,\mathrm{V}$. 
    (b) $V_{\mathrm{IN}}=12\,\mathrm{V}$. 
    (c) $V_{\mathrm{IN}}=24\,\mathrm{V}$.}
    \label{fig:TEM_On_Ti_2A}
\end{figure*}

\begin{figure*}[!t]
    \centering
    \makebox[\textwidth][c]{%
        \subfloat[]{
            \includegraphics[width=0.34\textwidth]{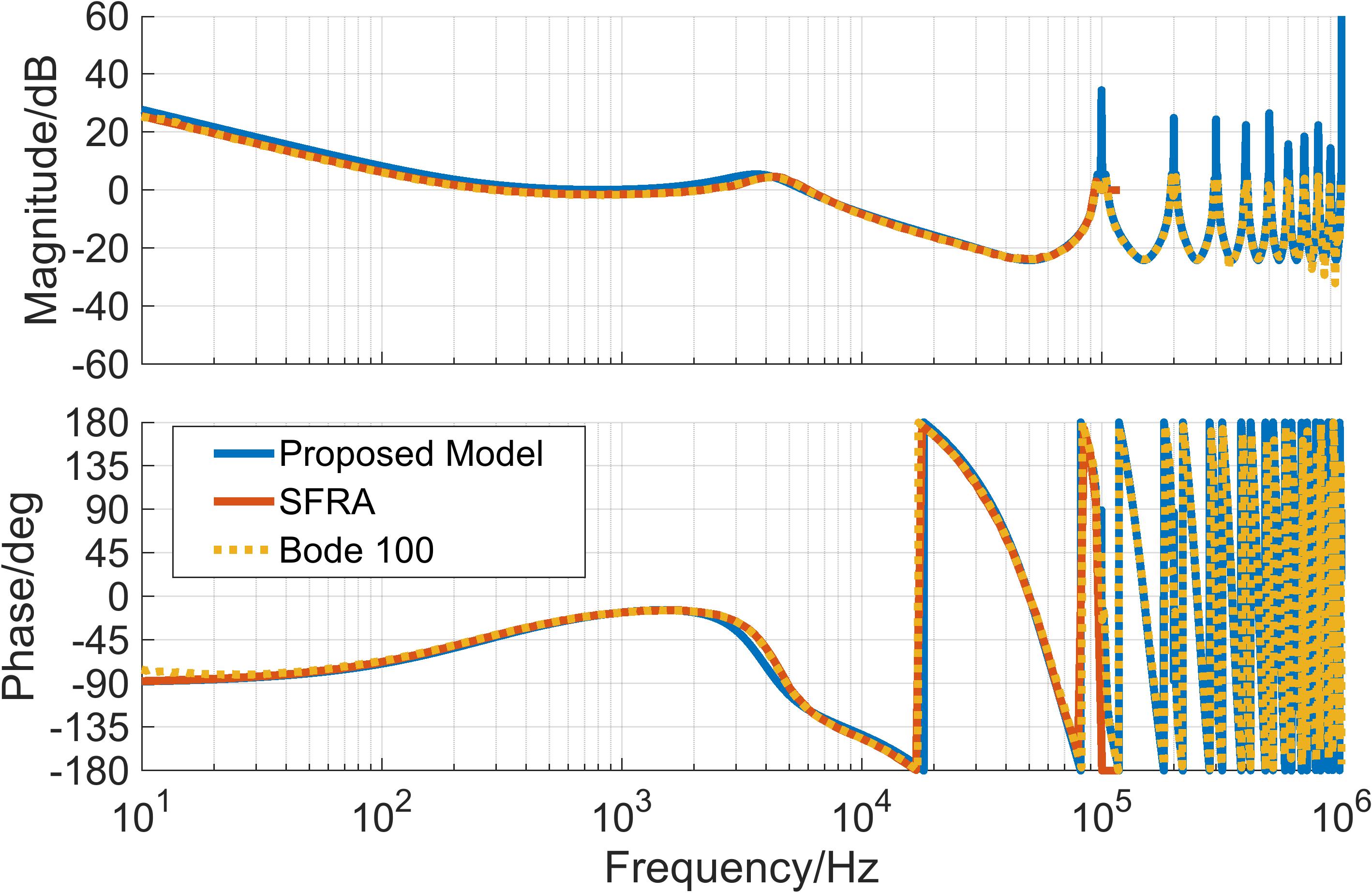}
            \label{fig:TEM_On_Tpul_5V_2A}
        }\hspace{-0.03\textwidth}
        \subfloat[]{
            \includegraphics[width=0.34\textwidth]{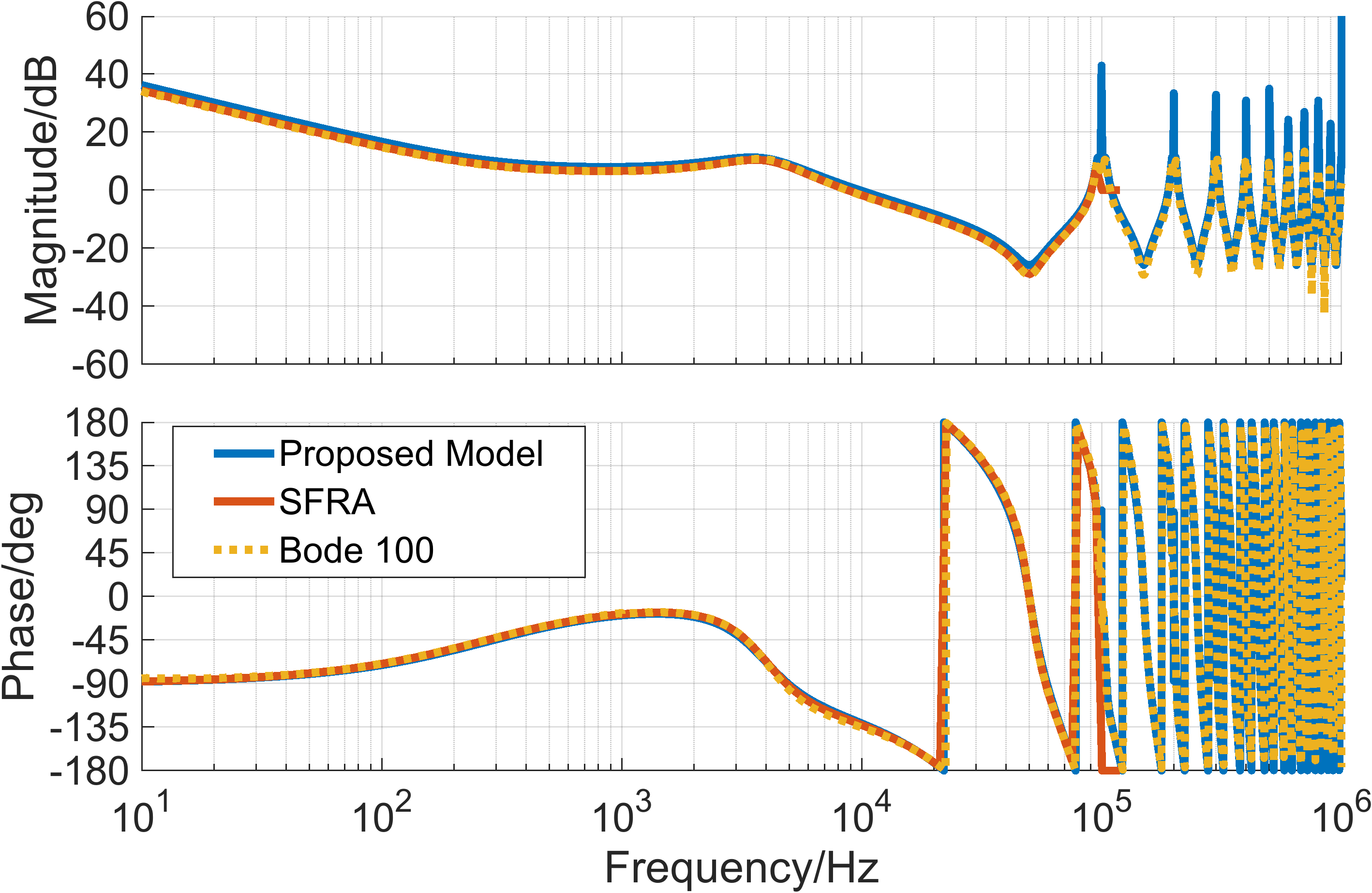}
            \label{fig:TEM_On_Tpul_12V_2A}
        }\hspace{-0.03\textwidth}
        \subfloat[]{
            \includegraphics[width=0.34\textwidth]{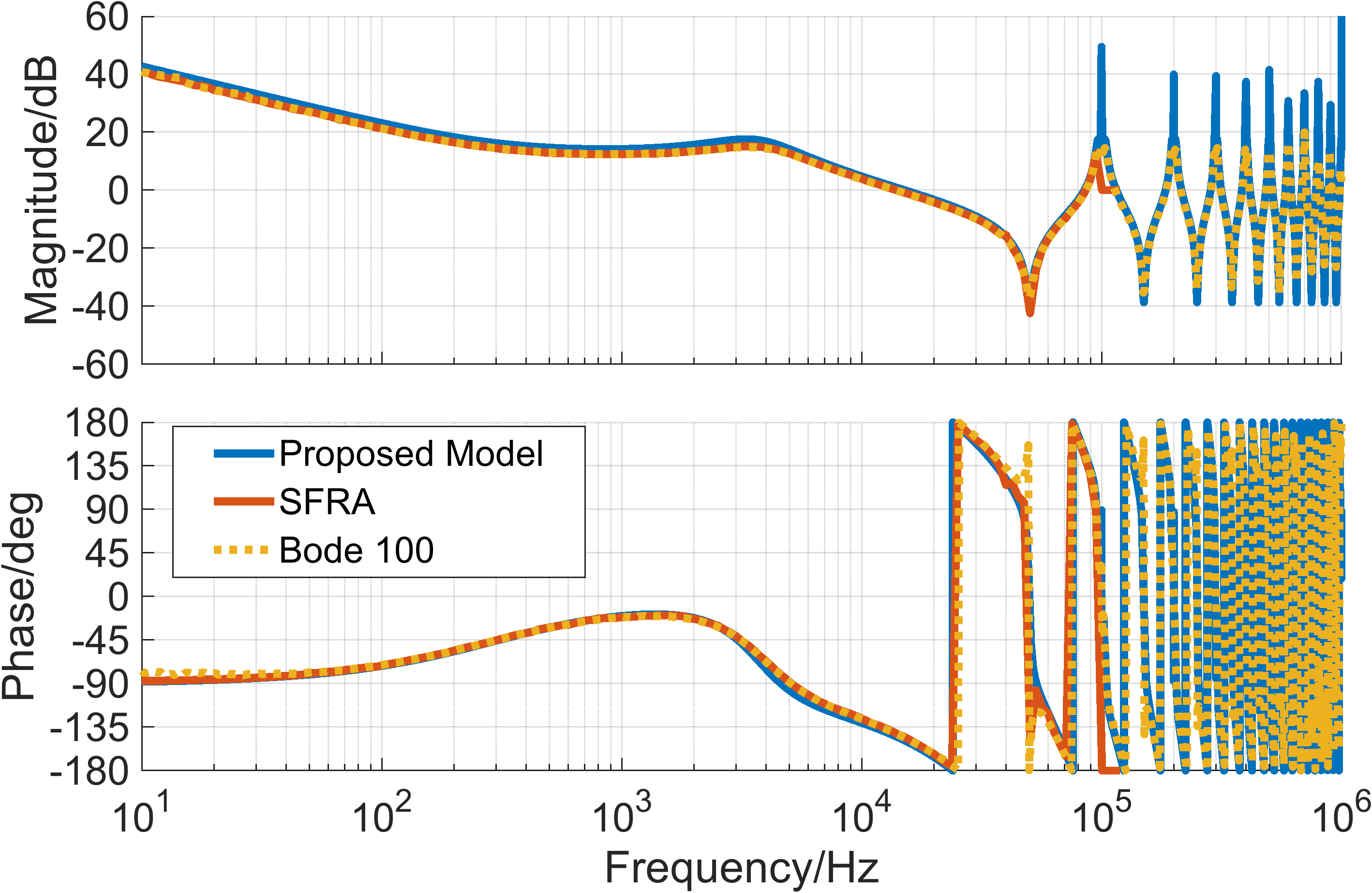}
            \label{fig:TEM_On_Tpul_24V_2A}
        }
    }

    \caption{Bode diagram of the calculated and measured digital loop gain $T_{pul}$, 
    (a) $V_{\mathrm{IN}}=5\,\mathrm{V}$. 
    (b) $V_{\mathrm{IN}}=12\,\mathrm{V}$. 
    (c) $V_{\mathrm{IN}}=24\,\mathrm{V}$.}
    \label{fig:TEM_On_Tpul_2A}
\end{figure*}

The converter operates under TEM with turn-on centered inductor current sampling. The circuit parameters, operating conditions, and design objectives are given in Table \ref{tab:parameters}. Using the design method in Section V.A, the PI parameters are determined to be $K_P=0.1592$ and $K_iT_S=0.047$. Fig.\ref{fig:TEM_On_Ti_12V_2A} and Fig.\ref{fig:TEM_On_Tpul_12V_2A} show that the measured $\omega_c$ and $\theta_c$ are $2\pi\times9.2\text{kHz}$ and $\pi/180^\circ\times51^\circ$, which validates the PI design method.

\begin{figure}[b]
    \centering
    \includegraphics[width=0.9\columnwidth]{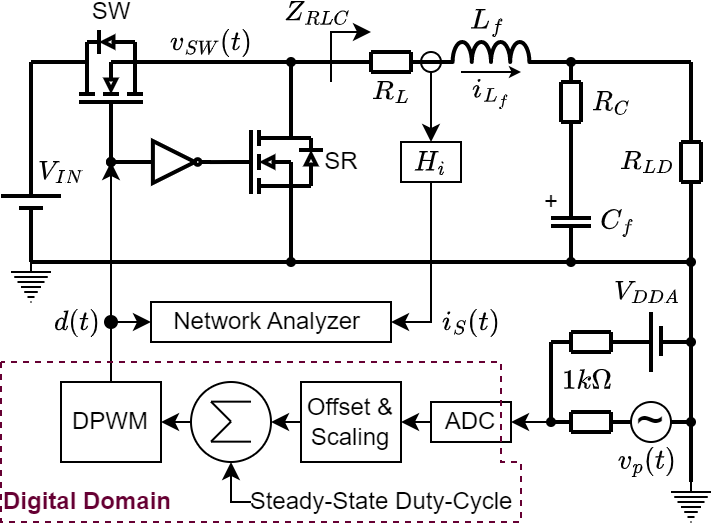}
    \caption{Plant characterization setup. $v_p(t)$ is the network analyzer AC output.}
    \label{fig:gid_ident}
\end{figure}

Fig.~\ref{fig:gid_ident} illustrates the open-loop experimental setup for identifying $G_{id}(s)H_i$. A perturbed PWM duty-ratio signal, $d(t)$, is generated by the DSP. Since the measurement is performed in open loop, as shown in Fig.~\ref{fig:gid_ident}, sideband effects do not influence the identified response. The Bode 100 network analyzer measures the transfer function from $d(t)$ to the sensed current $i_s(t)$, yielding:
\begin{equation}
    G_{id}(s)H_i = \frac{i_s(s)}{d(s)}
    = \frac{V_{IN} H_i}{V_D Z_{RLC}(s)},
\end{equation}
where $V_D$ denotes the amplitude of $d(t)$, and $Z_{RLC}(s)$ is the RLC impedance seen from the switching node $v_{SW}$. The identification is carried out on the actual hardware under $V_{IN}=12~\mathrm{V}$ and $D=0.25$. The measured response shows good agreement with the component values given in Table~\ref{tab:parameters}.

The $T_i$ measurement follows the conventional approach\cite{YanNaPartI}, in which a perturbation source (the injection transformer) is inserted in series with the ADC input and the network analyzer measures the response across it. Fig.~\ref{fig:TEM_On_Ti_2A} presents the measured $T_i$ at $I_{REF}=2~\mathrm{A}$ for various $V_{IN}$ and different duty cycles. The measurement result shows good agreement with the prediction.

Fig.~\ref{fig:TEM_On_Tpul_2A} compares the $T_\text{pul}$ response predicted by the proposed model with the measurements obtained using two different injection methods. In the first method, the perturbation signal generated by Bode 100 is sampled by the DSP, added to the PI controller output, and the pre- and post-injection PI controller outputs are routed back to the Bode 100 via DACs~\cite{Lin_Accurate_Digital_Buck,Ruan_Loopgain_Measure}. The second method implements a software-based frequency response analyzer (SFRA) directly on the DSP. In this approach, the SFRA injects discrete sinusoidal perturbations in software at selected frequencies and uses a discrete-time Fourier transform (DTFT) to extract the magnitude and phase response at each frequency. The process is repeated sequentially over the frequency range to construct the Bode plot. As shown in Fig.~\ref{fig:TEM_On_Tpul_2A}, the two measurement methods produce equivalent results and closely match the model predictions. In practice, SFRA is preferable because it does not require a dedicated network analyzer and can be easily embedded in the firmware.

\begin{figure}[t]
    \centering
    \includegraphics[width=0.9\columnwidth]{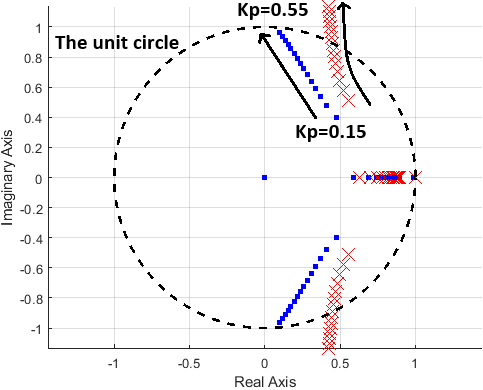}
    \caption{Closed-loop pole trajectories obtained from $1+T_\text{pul}$ as $K_P$ is swept from 0.15 to 0.55. The closed-loop poles move toward the outside of the unit circle as $K_P$ increases. The dots denote the results with DPWM-ADC synchronization included, whereas the crosses denote the results without DPWM-ADC synchronization.}
    \label{fig:KpScan}
\end{figure}

\begin{figure}[t]
    \centering
    \includegraphics[width=0.9\columnwidth]{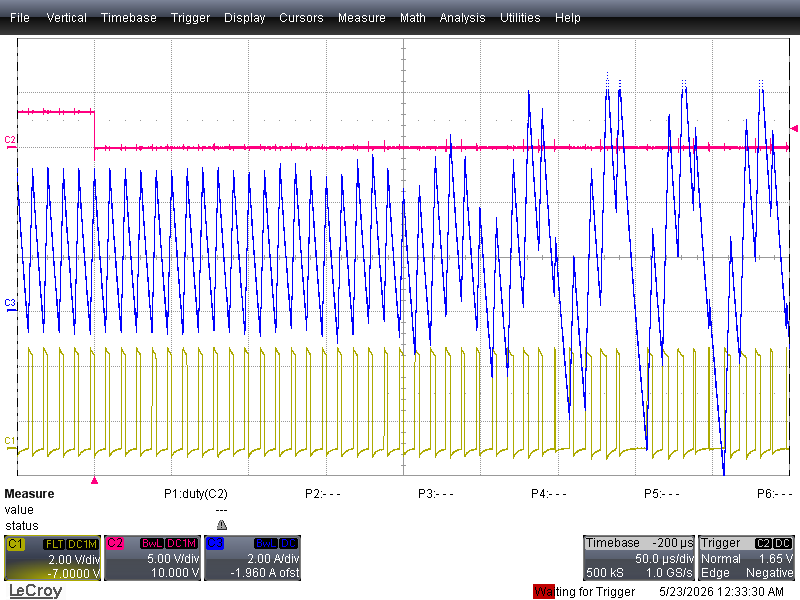}
    \caption{Experimental waveform for $K_P=0.55$. The converter starts up and reaches steady state with DPWM--ADC synchronization enabled, but becomes unstable after the synchronization is disabled. C1: PWM signal. C2: DPWM--ADC synchronization enable signal, a high level indicates that synchronization is enabled. C3: the inductor current $i_L$.}
    \label{fig:KpScan_Exp}
\end{figure}

According to Jury's criterion, a discrete-time closed-loop system is stable if and only if all closed-loop poles are inside the unit circle. Therefore, the closed-loop stability can be evaluated by examining the zeros of $1+T_\text{pul}$. Fig.~\ref{fig:KpScan} illustrates the closed-loop pole trajectories as $K_P$ swept from $0.15$ to $0.55$. With DPWM-ADC synchronization, the closed-loop poles remain inside the unit circle. By contrast, when DPWM-ADC synchronization is not applied, the system becomes unstable, even under the exact same steady-state operating condition and with the same PI controller. This is experimentally confirmed in Fig.~\ref{fig:KpScan_Exp}. The converter starts with $K_P=0.55$ and enters a stable steady-state with DPWM-ADC synchronization enabled. However, after disabling DPWM-ADC synchronization, i.e., by fixing the ADC sampling instant at its steady-state value, the converter starts to oscillate. These results confirm that the converter has substantially different small-signal characteristics depending on whether DPWM-ADC synchronization is enabled or not. Hence, the influence of DPWM-ADC synchronization must be accounted for in the stability analysis.

\section{Conclusions}
This paper has presented an accurate small-signal model that explicitly accounts for DPWM-ADC synchronization in digitally controlled Buck converters. Without introducing approximations in the derivation, the proposed model rigorously captures all sideband components based on sampled-data theory and provides accurate predictions of both analog and digital loop gains from DC to beyond the switching frequency. Three carrier modulations (TEM, LEM, and SM) have been analyzed. The use of modified Z-transform greatly simplifies the derivation of the digital loop gain.

A key finding is that DPWM-ADC synchronization introduces a feedthrough path that can significantly alter the small-signal behavior, contrary to the common assumption that its impact is negligible. Experimental results further demonstrate how such synchronization affects loop stability and verify the accuracy of the proposed model. In addition, a fully discrete-domain PI parameter design method has been developed and experimentally validated, enabling low-complexity compensator design and stability assessment. The proposed model can be reduced to \cite{YanNaPartII,Dragan,Lin_Accurate_Digital_Buck} by removing the feedthrough path introduced by DPWM-ADC synchronization.

The proposed model can be generalized to other converter topologies, such as the Boost converter. Future work will focus on extending the methodology to other converter topologies and variable-frequency modulations.


\bibliographystyle{Bibliography/IEEEtranTIE}
\bibliography{references}
	
\begin{IEEEbiography}[{\includegraphics[width=1in,height=1.25in,clip,keepaspectratio]{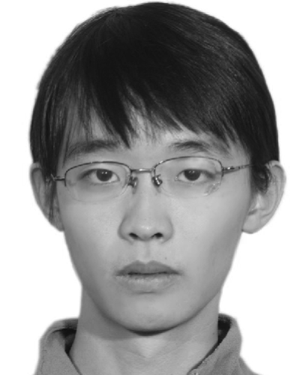}}]{Hang Zhou} (Member, IEEE) received the bachelor's and Ph.D. degrees in electrical engineering from the University of New South Wales, Sydney, NSW, Australia, in 2017 and 2022, respectively.

He is currently a Research Associate with the University of New South Wales in the field of power electronics, specializing in dc-dc converters. His research interests include low-voltage high-current dc-dc topologies, small-signal modeling of dc-dc converters, three-phase power factor correction, and high-voltage power sources.
\end{IEEEbiography}
\begin{IEEEbiography}[{\includegraphics[width=1in,height=1.25in,clip,keepaspectratio]{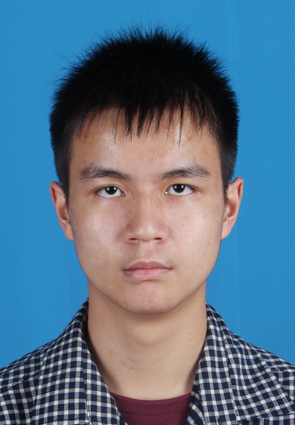}}]{Yuxin Yang} (Member, IEEE) was born in China. He received the Bachelor's degree in electrical engineering from the University of New South Wales (UNSW), Sydney, Australia, in 2024. He is currently pursuing the Master by Philosophy degree in electrical engineering at UNSW. His major field of research is power electronics and control systems.

He is currently conducting research on small-signal modeling of power electronic systems, sampled-data control theory, and the foundational theory of control systems. His current research focuses on unifying sampled-data models with continuous-time system representations using rigorous mathematical tools.
\end{IEEEbiography}
\begin{IEEEbiography}[{\includegraphics[width=1in,height=1.25in,clip,keepaspectratio]{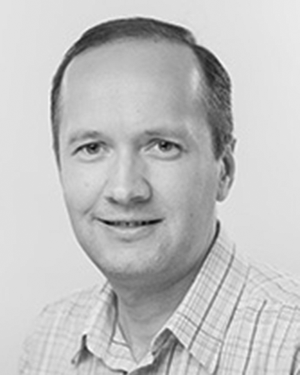}}]{Branislav Hredzak}  (Senior Member, IEEE) received the Ing. degree in electrical engineering from the Technical University of Kosice, Kosice, Slovak Republic, in 1993, and the Ph.D. degree in electrical engineering from the Napier University of Edinburgh, Edinburgh, U.K., in 1997. He was a Lecturer and a senior Researcher in Singapore from 1997 to 2007. He is currently an Associate Professor with the School of Electrical Engineering and Telecommunications, UNSW, Sydney, NSW, Australia. His research interests include the control of distributed renewable energy sources, hybrid and reconfigurable energy storage technologies, virtual power plants, and advanced control systems for power converters and energy storage systems.
\end{IEEEbiography}
\begin{IEEEbiography}[{\includegraphics[width=1in,height=1.25in,clip,keepaspectratio]{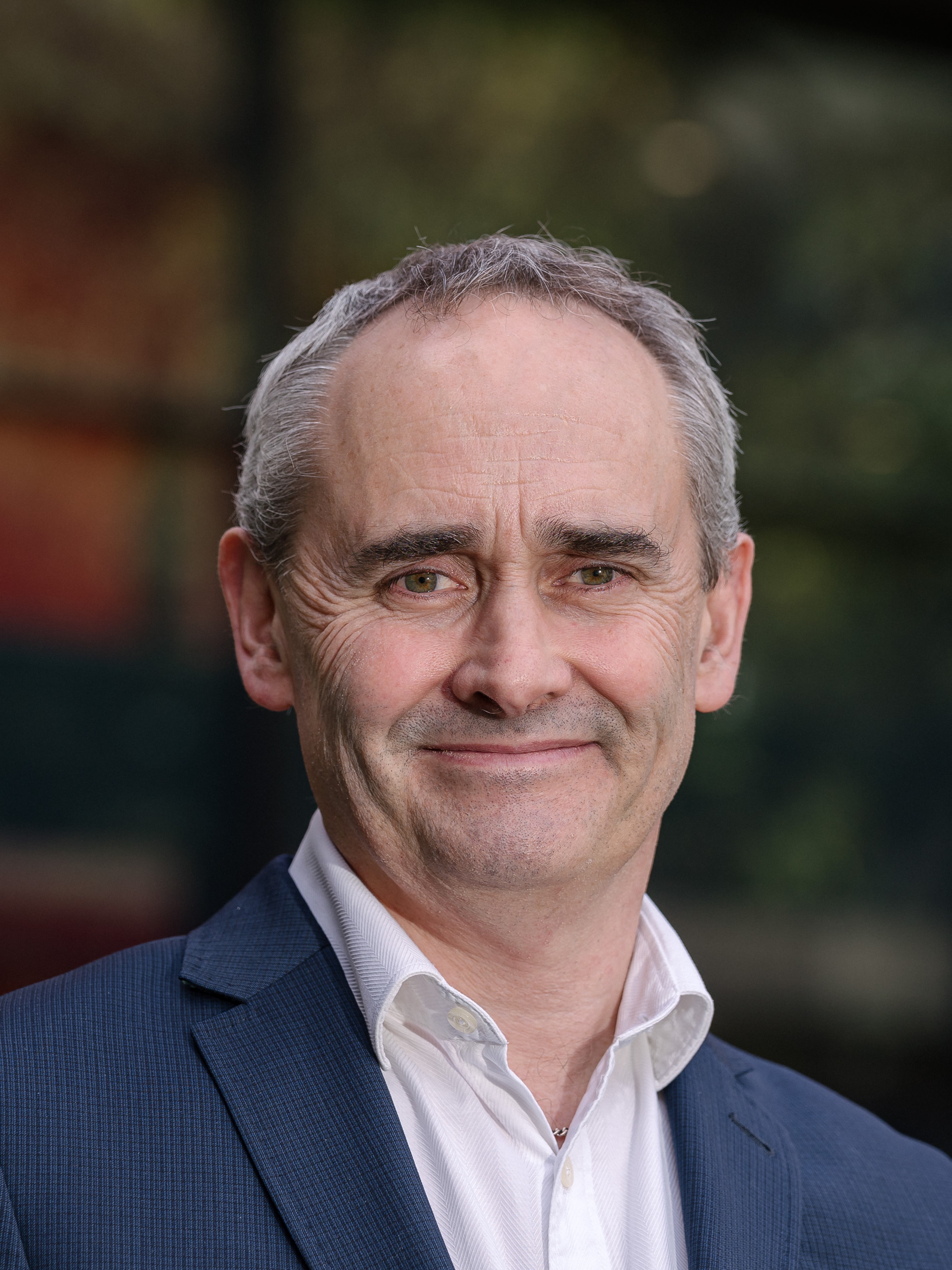}}]{John Edward Fletcher} (Senior Member, IEEE) received the B.Eng. (first-class Hons.) and Ph.D. degrees in electrical and electronic engineering from Heriot-Watt University, Edinburgh, U.K., in 1991 and 1995, respectively. He is currently a Professor with the University of New South Wales, Sydney, NSW, Australia. His research interests include power electronics, drives, and energy conversion. Dr. Fletcher is a Chartered Engineer in the U.K., and Fellow of the Institution of Engineering and Technology.
\end{IEEEbiography}

\end{document}